\documentclass[aps, prl, twocolumn, superscriptaddress,preprintnumbers]{revtex4-2}
\usepackage{bm, amsmath, amsfonts, amssymb, ascmac, mathtools, braket}
\usepackage{times}
\usepackage{multirow}
\usepackage{graphicx}
\usepackage{float, color, xcolor}

\usepackage[whole]{bxcjkjatype} 
\usepackage{subfigure} 
\usepackage{amscd} 

\usepackage{bbm}
\usepackage{tabularx}

\usepackage{comment}

\usepackage[
pagebackref=false,
colorlinks=true,
linkcolor=blue,
urlcolor=blue,
filecolor=black,
citecolor=red,
pdfstartview=FitV,
pdftitle={},
pdfauthor={},
pdfsubject={},
pdfkeywords={},
pdfpagemode=None,
bookmarksopen=true
]{hyperref}

\newcommand{\ii}{\text{i}}

\newcommand{\bk}{{\bm{k}}}

\newcommand{\Z}{{\mathbb{Z}}}

\begin{document}

\title{Non-Hermitian Origin of Detachable Boundary States in Topological Insulators}

\author{Daichi Nakamura}
\email{daichi.nakamura@issp.u-tokyo.ac.jp}
\affiliation{Institute for Solid State Physics, University of Tokyo, Kashiwa, Chiba 277-8581, Japan}

\author{Ken Shiozaki}
\email{ken.shiozaki@yukawa.kyoto-u.ac.jp}
\affiliation{Center for Gravitational Physics and Quantum Information, Yukawa Institute for Theoretical Physics, Kyoto University, Kyoto 606-8502, Japan}

\author{Kenji Shimomura}
\affiliation{Center for Gravitational Physics and Quantum Information, Yukawa Institute for Theoretical Physics, Kyoto University, Kyoto 606-8502, Japan}

\author{Masatoshi Sato}
\affiliation{Center for Gravitational Physics and Quantum Information, Yukawa Institute for Theoretical Physics, Kyoto University, Kyoto 606-8502, Japan}

\author{Kohei Kawabata}
\email{kawabata@issp.u-tokyo.ac.jp}
\affiliation{Institute for Solid State Physics, University of Tokyo, Kashiwa, Chiba 277-8581, Japan}

\date{\today}
\preprint{YITP-24-82}

\begin{abstract}
While topology can impose obstructions to exponentially localized Wannier functions, certain topological insulators are exempt from such Wannier obstructions. 
The absence of the Wannier obstructions can further accompany topological boundary states that are detachable from the bulk bands.
Here, we elucidate a close connection between these detachable topological boundary states and non-Hermitian topology.
Identifying Hermitian topological boundary states as non-Hermitian topology, we demonstrate that intrinsic non-Hermitian topology leads to the inevitable spectral flow.
By contrast, we show that extrinsic non-Hermitian topology underlies the detachment of topological boundary states and clarify anti-Hermitian topology of the detached boundary states.
Based on this connection and $K$-theory, we complete the tenfold classification of Wannier localizability and detachable topological boundary states.
\end{abstract}

\maketitle

Topological insulators and superconductors play a pivotal role in modern condensed matter physics~\cite{HK-review, QZ-review}, 
generally classified by the fundamental tenfold internal symmetry classification~\cite{AZ-97, Schnyder-08, *Ryu-10, Kitaev-09, CTSR-review}.
A distinctive feature of the topologically nontrivial bulk bands is the emergence of anomalous states at boundaries.
Additionally, certain topology, such as the Chern number, imposes obstructions to constructing exponentially localized Wannier functions in the bulk bands~\cite{Thouless-84, Brouder-07, Soluyanov-11, Read-17, Vanderbilt-textbook}.
The Wannier localizability also provides a foundation of topological phases, including topological crystalline insulators~\cite{Po-18}.
Conversely, other types of topology do not necessarily lead to the Wannier obstructions~\cite{Kohn-59, Kivelson-82, Ono-20}.
Remarkably, contrary to the conventional intuition of the bulk-boundary correspondence, such Wannier-localizable topological insulators have recently been shown to host boundary states that are detachable from the bulk bands~\cite{Alexandradinata-21, Altland-24}.
In this Letter, we reveal an intimate connection of these detachable topological boundary states with non-Hermitian topology.

Recently, topological characterization of non-Hermitian systems has attracted considerable interest~\cite{Rudner-09, Sato-11, *Esaki-11, Hu-11, Schomerus-13, Longhi-15, Lee-16, Leykam-17, Xu-17, Shen-18, *Kozii-17, Takata-18, MartinezAlvarez-18, Gong-18, *Kawabata-19, YW-18-SSH, *YSW-18-Chern, Kunst-18, McDonald-18, Lee-Thomale-19, Liu-19, Lee-Li-Gong-19, KSUS-19, ZL-19, Herviou-19, Zirnstein-19, Borgnia-19, KBS-19, Yokomizo-19, JYLee-19, Wanjura-20, Zhang-20, OKSS-20, Bessho-21, Denner-21, Okugawa-20, KSS-20, KSR-21, Zhang-22, Sun-21, Franca-22, Nakamura-24, Wang-24, Nakai-24, Ma-24, Schindler-23, Nakamura-23, Hamanaka-24, BBK-review, Okuma-Sato-review}.
Non-Hermiticity arises from exchange of energy and particles with the environment~\cite{Konotop-review, Christodoulides-review}, and yields various topological phenomena unique to open systems~\cite{Poli-15, Zeuner-15, Zhen-15, *Zhou-18, Weimann-17, Xiao-17, St-Jean-17, Bahari-17, Zhao-18, Bandres-18, Zhao-19, Ghatak-19-skin-exp, Helbig-19-skin-exp, Xiao-19-skin-exp, Weidemann-20-skin-exp, Wengang-21, Liang-22}.
Importantly, these non-Hermitian topological phenomena stem from two types of complex-energy gaps: 
point and line gaps~\cite{KSUS-19}.
Non-Hermitian systems with point gaps are continuously deformable to unitary systems and thus can be intrinsic to non-Hermitian systems.
Such intrinsic non-Hermitian topology gives rise to the skin effect~\cite{Lee-16, YW-18-SSH, Kunst-18, Yokomizo-19, Zhang-20, OKSS-20} and exceptional points~\cite{Shen-18, KBS-19, Denner-21, Nakamura-24}.
Conversely, non-Hermitian systems with real (imaginary) line gaps are continuously deformable to Hermitian (anti-Hermitian) systems and therefore have counterparts in conventional Hermitian systems.
The interplay of point and line gaps enriches the topological classification of non-Hermitian systems based on the 38-fold internal symmetry classes~\cite{KSUS-19}.

In this Letter, we uncover a hidden relationship between non-Hermitian topology and Hermitian detachable topological boundary states.
Associating topological boundary states with point-gap topology, we demonstrate that intrinsic and extrinsic point-gap topology results in nondetachable and detachable topological boundary states, respectively.
Utilizing this connection and $K$-theory, we complete the tenfold classification of Wannier localizability and detachable topological boundary states (Table~\ref{tab: classification}).
We also elucidate topology of the detached boundary states, generally classified in Table~\ref{tab: topo inv}.

\renewcommand{\arraystretch}{1.0}

\begin{table*}[t]
	\centering
	\caption{Wannier localizability of the $d$-dimensional bulk and detachability of $\left( d-1 \right)$-dimensional boundary states in topological insulators and superconductors.
    The tenfold Altland-Zirnbauer symmetry classes comprise time-reversal symmetry (TRS), particle-hole symmetry (PHS), and chiral symmetry (CS).
    The entries with ``$\checkmark$" (``$\times$") accompany Wannier nonlocalizable (localizable) topological phases in the bulk and exhibit nondetachable (detachable) boundary states with intrinsic (extrinsic) non-Hermitian topology.
    The entries with ``$\checkmark/\times$" specify the Wannier nonlocalizable and localizable topological phases for the odd and even numbers of topological invariants, respectively.}
	\label{tab: classification}
     \begin{tabular}{c|ccc|cccccccc} \hline \hline
    ~~Class~~ & ~~TRS~~ & ~~PHS~~ & ~~CS~~ & ~~$d=1$~~ & ~~$d=2$~~ & ~~$d=3$~~ & ~~$d=4$~~ & ~~$d=5$~~ & ~~$d=6$~~ & ~~$d=7$~~ & ~~$d=8$~~ \\ \hline
    A & $0$ & $0$ & $0$ & $0$ & $\mathbb{Z}^{\checkmark}$ & $0$ & $\mathbb{Z}^{\checkmark}$ & $0$ & $\mathbb{Z}^{\checkmark}$ & $0$ & $\mathbb{Z}^{\checkmark}$ \\
    AIII & $0$ & $0$ & $1$ & $\mathbb{Z}^{\times}$ & $0$ & $\mathbb{Z}^{\times}$ & $0$ & $\mathbb{Z}^{\times}$ & $0$ & $\mathbb{Z}^{\times}$ & $0$ \\ \hline
    AI & $+1$ & $0$ & $0$ & $0$ & $0$ & $0$ & $2\mathbb{Z}^{\checkmark}$ & $0$ & $\mathbb{Z}_2^{\checkmark}$ & $\mathbb{Z}_2^{\checkmark}$ & $\mathbb{Z}^{\checkmark}$ \\
    BDI & $+1$ & $+1$ & $1$ & $\mathbb{Z}^{\times}$ & $0$ & $0$ & $0$ & $2\mathbb{Z}^{\times}$ & $0$ & $\mathbb{Z}_2^{\times}$ & $\mathbb{Z}_2^{\times}$ \\
    D & $0$ & $+1$ & $0$ & $\mathbb{Z}_2^{\times}$ & $\mathbb{Z}^{\checkmark}$ & $0$ & $0$ & $0$ & $2\mathbb{Z}^{\checkmark}$ & $0$ & $\mathbb{Z}_2^{\times}$ \\
    DIII & $-1$ & $+1$ & $1$ & $\mathbb{Z}_2^{\times}$ & $\mathbb{Z}_2^{\checkmark}$ & $\mathbb{Z}^{\checkmark/\times}$ & $0$ & $0$ & $0$ & $2\mathbb{Z}^{\times}$ & $0$ \\
    AII & $-1$ & $0$ & $0$ & $0$ & $\mathbb{Z}_2^{\checkmark}$ & $\mathbb{Z}_2^{\checkmark}$ & $\mathbb{Z}^{\checkmark}$ & $0$ & $0$ & $0$ & $2\mathbb{Z}^{\checkmark}$ \\
    CII & $-1$ & $-1$ & $1$ & $2\mathbb{Z}^{\times}$ & $0$ & $\mathbb{Z}_2^{\times}$ & $\mathbb{Z}_2^{\times}$ & $\mathbb{Z}^{\times}$ & $0$ & $0$ & $0$ \\
    C & $0$ & $-1$ & $0$ & $0$ & $2\mathbb{Z}^{\checkmark}$ & $0$ & $\mathbb{Z}_2^{\times}$ & $\mathbb{Z}_2^{\times}$ & $\mathbb{Z}^{\checkmark}$ & $0$ & $0$ \\
    CI & $+1$ & $-1$ & $1$ & $0$ & $0$ & $2\mathbb{Z}^{\times}$ & $0$ & $\mathbb{Z}_2^{\times}$ & $\mathbb{Z}_2^{\checkmark}$ & $\mathbb{Z}^{\checkmark/\times}$ & $0$ \\ \hline \hline
  \end{tabular}
\end{table*}

\renewcommand{\arraystretch}{1}

\textit{Non-Hermitian topology}.---A distinctive feature of non-Hermitian systems is complex-valued spectra, leading to point and line gaps~\cite{Gong-18, KSUS-19}.
In the presence of a point (line) gap, complex-energy bands do not cross a reference point (line) in the complex-energy plane.
Specifically, a non-Hermitian Hamiltonian $H$ is defined to have a point gap for $E_n \neq E_{\rm P}$, where $E_n$'s are complex eigenenergies of $H$, and $E_{\rm P} \in \mathbb{C}$ is a reference energy.
By contrast, $H$ is defined to have a real (imaginary) line gap for $\mathrm{Re}\,E_n \neq E_{\rm L}$ ($\mathrm{Im}\,E_n \neq E_{\rm L}$) with $E_{\rm L} \in \mathbb{R}$, implying separable bands~\cite{Shen-18}.
Even if a point gap is open, a line gap is not necessarily open. 
However, a point gap is always open when a line gap is open by placing a reference point on the reference line.
Consequently, some point-gap topology is continuously deformable to line-gap topology and has a counterpart in Hermitian or anti-Hermitian systems.
In contrast to such extrinsic non-Hermitian topology, certain point-gap topology is irreducible to line-gap topology and thus intrinsic to non-Hermitian systems, serving as the origin of the skin effect~\cite{Zhang-20, OKSS-20}.

\begin{figure}[tbp]
    \centering
    \includegraphics[width=\linewidth]{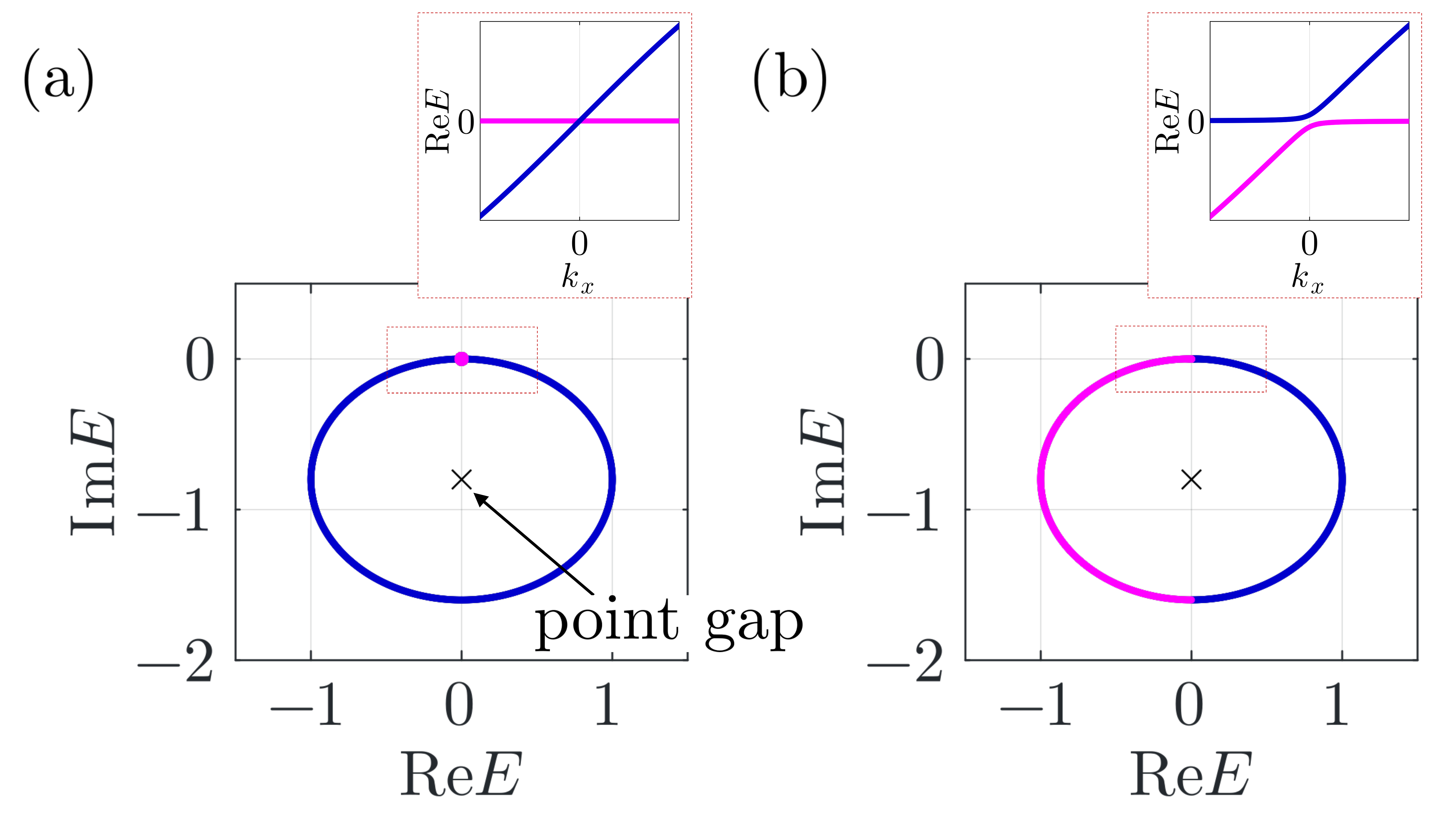} 
    \caption{Intrinsic non-Hermitian topology of the chiral edge state.
    Complex spectrum of the point-gapped non-Hermitian model $\tilde{H}_{\rm A} \left( k_x \right)$ in Eq.~(\ref{eq: 1D class A perturbation}) ($\gamma = 0.8$) for (a)~$v = 0$ and (b)~$v=0.03$.
    (a)~Around $E = 0$, $\tilde{H}_{\rm A} \left( k \right)$ (main panel) reduces to the chiral edge state $E \left( k_x \right) \simeq k_x$ (inset).
    (b)~The perturbation $v \neq 0$ leads to a swap of two complex bands (blue and magenta curves), protecting the spectral flow.}
        \label{fig: 2D class A}
\end{figure}

\textit{Chiral edge state and intrinsic non-Hermitian topology}.---To illustrate the role of intrinsic non-Hermitian topology in nondetachable topological boundary
states, we investigate a chiral edge state in a two-dimensional Chern insulator: 
$\mathcal{H}_{\rm bdy} \left( k_x \right) = k_x$.
In the adiabatic change of momentum, the chiral edge state exhibits spectral flow from negative infinity to positive infinity, inevitably requiring the attachment to the bulk degrees of freedom.

We identify the chiral edge state with a point-gapped non-Hermitian system [Fig.~\ref{fig: 2D class A}\,(a)],
\begin{equation}
    H_{\rm A} \left( k_x \right) = \sin k_x + \ii \gamma \left( \cos k_x - 1 \right),
        \label{eq: 1D A}
\end{equation}
where $\gamma > 0$ is the degree of non-Hermiticity and describes the nonreciprocal hopping in real space~\cite{Hatano-Nelson-96, *Hatano-Nelson-97}.
Hereafter, we use $\mathcal{H}$ ($H$) for Hermitian (non-Hermitian) Hamiltonians.
This identification is formalized by adding dissipation to the edges~\cite{Ma-24, Schindler-23, Nakamura-23} or incorporating the coupling between the bulk and edges~\cite{Hamanaka-24}, where the bulk degrees of freedom at infinity correspond to decaying eigenstates with $\mathrm{Im}\,E < 0$.
Around the Fermi surface $\mathrm{Re}\,E=0$, this non-Hermitian system $H_{\rm A} \left( k_x \right)$ hosts two modes $k_x$ and $-k_x - 2\ii \gamma$, only the former of which survives since the latter decays with a finite lifetime $1/2\gamma$~\cite{JYLee-19, Bessho-21}.

This identification is generally guaranteed by point-gap topology~\cite{Gong-18, KSUS-19}, i.e.,
\begin{equation}
    W_1 \left[ H \right]
    \coloneqq -\oint_{\rm BZ} \frac{dk}{2\pi\ii} \left( \frac{d}{dk} \log \det \left[ H \left( k \right) - E_{\rm P} \right] \right).
\end{equation}
Specifically, $H_{\rm A} \left( k_x \right)$ in Eq.~(\ref{eq: 1D A}) is characterized by $W_1 \left[ H_{\rm A} \right] = 1$ for the reference energy $E_{\rm P}$ inside the loop of the complex spectrum.
This complex-spectral winding always vanishes for Hermitian systems and is intrinsic to non-Hermitian systems.
Thus, the point gap cannot be opened or reduced to any line gaps while preserving the point-gap topology.
Consequently, the chiral edge state around $E = 0$ cannot be detached from other eigenstates with $\mathrm{Im}\,E < 0$ that play a role of the bulk degree of freedom.
To further confirm this stability, we couple the chiral edge state to a trivial band $H = 0$ [Fig.~\ref{fig: 2D class A}\,(b)], 
\begin{equation}   
    \tilde{H}_{\rm A} \left( k_x \right) = \begin{pmatrix}
        H_{\rm A} \left( k_x \right) & v \\
        v^{*} & 0
    \end{pmatrix},
        \label{eq: 1D class A perturbation}
\end{equation}
with the coupling amplitude $v \in \mathbb{C}$.
As momentum is traversed through the Brillouin zone, the two bands swap with each other, prohibiting the point-gap opening~\cite{supplement}.

\textit{Dirac surface state and extrinsic non-Hermitian topology}.---In contrast to the chiral edge states, we demonstrate that the detachable topological boundary states accompany extrinsic non-Hermitian topology.
As a prime example, we investigate a Dirac surface state of a three-dimensional topological insulator protected solely by chiral symmetry:
$\mathcal{H}_{\rm bdy} \left( \bm{k} \right) = k_x \sigma_x + k_y \sigma_y$.
This Dirac Hamiltonian indeed respects chiral symmetry $\Gamma \mathcal{H}_{\rm bdy} \left( \bm{k} \right) \Gamma^{-1} = - \mathcal{H}_{\rm bdy} \left( \bm{k} \right)$ with $\Gamma = \sigma_z$ and hence belongs to class AIII~\cite{CTSR-review}.
Since chiral symmetry is relevant only to zero energy, $\mathcal{H}_{\rm bdy} \left( \bm{k} \right)$ accompanies no spectral flow and is detachable from the bulk bands~\cite{Altland-24, supplement}.

\begin{figure}[tbp]
    \centering
    \includegraphics[width=\linewidth]{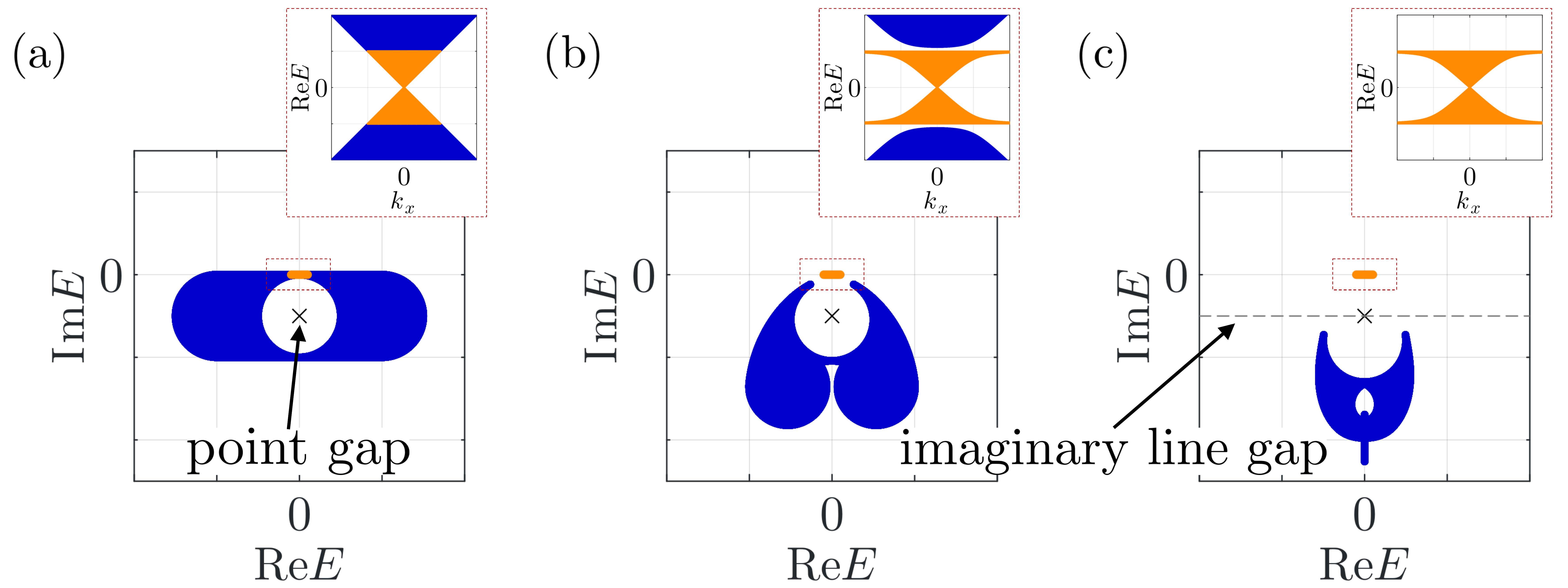} 
    \caption{Detachment of topological boundary states in three-dimensional topological insulators protected by chiral symmetry (class AIII).
    (a)~The Dirac surface state (inset) is identified as a chiral-symmetric non-Hermitian system with point-gap topology (main panel).
    (b, c)~Owing to the extrinsic nature of the point-gap topology, chiral-symmetry-preserving perturbations open an imaginary line gap (dashed line) while preserving the point gap.
    The Dirac surface state around $E = 0$ is separated from the other decaying bands with $\mathrm{Im}\, E < 0$, corresponding to the detachment from the bulk bands.}
        \label{fig: 3D class AIII}
\end{figure}

We identify this Dirac surface state as a point-gapped non-Hermitian system in two dimensions [Fig.~\ref{fig: 3D class AIII}\,(a)], 
\begin{align}
    H_{\rm AIII} \left( \bm{k} \right) &= \left( \sin k_x \right) \sigma_x + \left( 1- \cos k_x -\cos k_y \right) \sigma_y \nonumber \\
    &\qquad\qquad\qquad\quad + \ii \gamma \left( \sin k_y - 1 \right),
    \label{eq: 2dAIII}
\end{align}
with $\gamma > 0$.
This non-Hermitian Dirac model also respects chiral symmetry~\cite{KSUS-19}:
\begin{equation}
    \Gamma H_{\rm AIII}^{\dag} \left( \bm{k} \right) \Gamma^{-1} = - H_{\rm AIII} \left( \bm{k} \right)
        \label{eq: CS}
\end{equation}
with $\Gamma = \sigma_z$.
Around the Fermi surface $\mathrm{Re}\,E = 0$ (i.e., $k_x = 0$, $k_y = \pm \pi/2$), it comprises two modes, $k_x \sigma_x + k_y \sigma_y$ and $k_x \sigma_x - k_y \sigma_y -2\ii \gamma$, only the former of which survives and reduces to $\mathcal{H}_{\rm bdy} \left( \bm{k} \right)$~\cite{JYLee-19, Bessho-21}.

\begin{figure}[tbp]
    \centering
    \includegraphics[width=\linewidth]{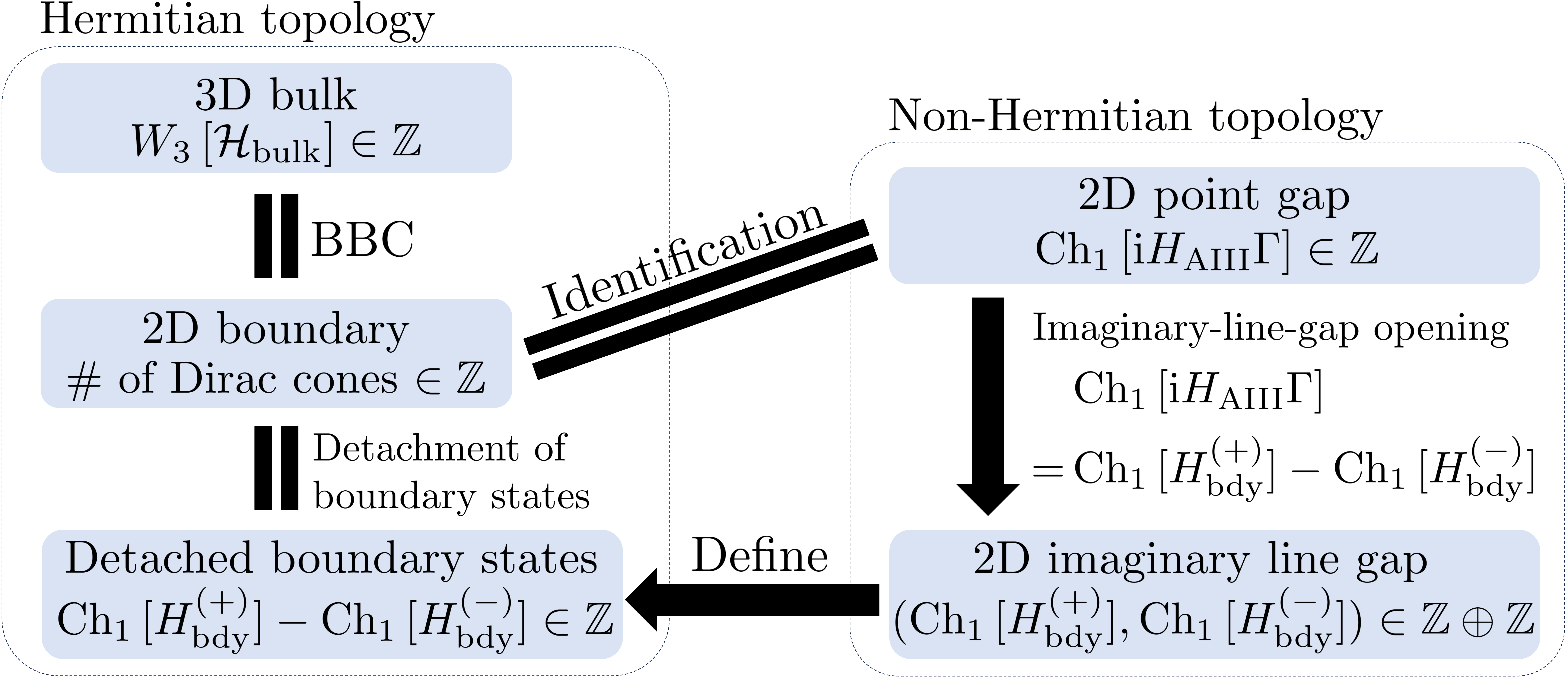}
    \caption{Relationship between Hermitian and non-Hermitian topology for three-dimensional (3D) class AIII. 
    Based on the bulk-boundary correspondence (BBC) and the identification of Hermitian topological boundary states with non-Hermitian point-gapped systems, all the topological invariants connected by the equal signs coincide. 
    Furthermore, if imaginary line gaps open, the corresponding imaginary-line-gap topology defines topological invariants of detached boundary states.}
        \label{fig: topology relation}
\end{figure}

Point-gap topology of this chiral-symmetric non-Hermitian Dirac model is captured by the (first) Chern number of the Hermitian matrix $\ii \left[ H_{\rm AIII} \left( \bm{k} \right) - E_{\rm P} \right] \Gamma$, where the reference point $E_{\rm P}$ is chosen to respect chiral symmetry (i.e., $E_{\rm P} \in \ii \mathbb{R}$)~\cite{KSUS-19}.
From the bulk-boundary correspondence, this boundary point-gap topology $\mathrm{Ch}_1 \left[ \ii H_{\rm AIII} \Gamma\right]$ further coincides with the three-dimensional winding number $W_3 \left[ \mathcal{H}_{\rm bulk} \right]$ of the original Hermitian bulk $\mathcal{H}_{\rm bulk}$ (see Fig.~\ref{fig: topology relation}):
\begin{equation}
    \mathrm{Ch}_1 \left[ \ii H_{\rm AIII} 
    \Gamma\right] = W_3 \left[ \mathcal{H}_{\rm bulk} \right].
        \label{eq: AIII BBC}
\end{equation}
A crucial distinction from the chiral edge states is that this point-gap topology is continuously deformable to imaginary-line-gap topology while preserving the point gap and chiral symmetry (Fig.~\ref{fig: 3D class AIII}).
When we couple $H_{\rm AIII} \left( \bm{k} \right)$ to a trivial band $H=\varepsilon \sigma_x$ ($\varepsilon \geq 0$) by
\begin{equation}
    \tilde{H}_{\rm AIII} \left( \bm{k} \right) = \begin{pmatrix}
        H_{\rm AIII} \left( \bm{k} \right) & v \sigma_- \\
        v^{*}\sigma_+ & \varepsilon \sigma_x
    \end{pmatrix}
        \label{eq: 2D class AIII perturbation}
\end{equation}
with $v \in \mathbb{C}$, the bands around $E = 0$ are separated from the other bands with $\mathrm{Im}\,E < 0$, leading to the opening of an imaginary line gap~\cite{supplement}.
As discussed above, decaying eigenstates with $\mathrm{Im}\,E < 0$ describe the bulk degrees of freedom.
Thus, the imaginary-line-gap opening indicates the detachment of the Dirac surface state from the bulk bands.

Once the boundary states are detached, we can introduce a (non-Hermitian) Hamiltonian $H_{\rm bdy} \left( \bm{k} \right)$ projected onto the detached boundary bands [see the red dotted rectangle in Fig.~\ref{fig: 3D class AIII}\,(c)], and define its imaginary-line-gap topology.
Notably, chiral symmetry in Eq.~(\ref{eq: CS}) behaves like $\mathbb{Z}_2$ unitary symmetry that commutes with $H_{\rm bdy} \left( \bm{k} \right)$~\cite{pH}.
Hence, $H_{\rm bdy} \left( \bm{k} \right)$ hosts two independent Chern numbers $\mathrm{Ch}_1\,[ H_{\rm bdy}^{(\pm)} ]$ in the subspaces of $\Gamma = \pm 1$. 
This $\mathbb{Z} \oplus \mathbb{Z}$ topological classification for imaginary line gaps includes the $\mathbb{Z}$ classification for point gaps, showing the extrinsic nature of the point-gap topology.
A map from $\mathbb{Z} \oplus \mathbb{Z}$ to $\mathbb{Z}$ is given as~\cite{OKSS-20, Shiozaki, supplement}
\begin{equation}
    \mathrm{Ch}_1\,[ H_{\rm bdy}^{(+)} ] - \mathrm{Ch}_1\,[ H_{\rm bdy}^{(-)} ] = \mathrm{Ch}_1 \left[ \ii H_{\rm AIII} \Gamma \right],
\end{equation}
where a reference point and line are chosen appropriately.
Thus, in combination with Eq.~(\ref{eq: AIII BBC}), as a direct consequence of the bulk topology $W_3 \left[ \mathcal{H}_{\rm bulk} \right] \neq 0$, the detached boundary states inevitably exhibit the nontrivial Chern number $\mathrm{Ch}_1\,[ H_{\rm bdy}^{(+)} ] \neq 0$ or $\mathrm{Ch}_1\,[ H_{\rm bdy}^{(-)} ] \neq 0$, 
consistent with the surface Chern number introduced in Refs.~\cite{Alexandradinata-21, Altland-24}. 
We summarize the connection between these topological invariants in Fig.~\ref{fig: topology relation}.

\begin{figure}[t]
    \centering
    \includegraphics[width=1.0\linewidth]{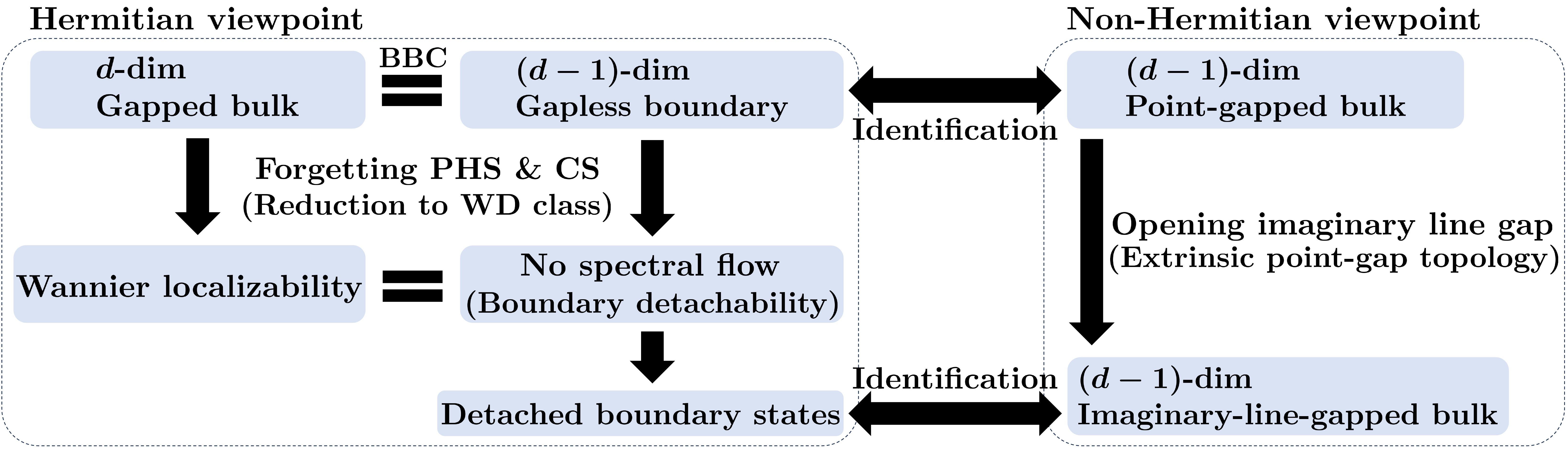} 
    \caption{Bulk Wannier localizability and detachable topological boundary states.
    ``BBC" denotes bulk-boundary correspondence, ``PHS" and ``CS" particle-hole and chiral symmetries, respectively, and ``WD" Wigner-Dyson.}
        \label{fig: commutative diagram}
\end{figure}

\textit{Classification}.---We extend our formulation to generic topological materials and classify
Wannier localizability and detachable topological boundary states in Table~\ref{tab: classification} (see also Fig.~\ref{fig: commutative diagram}).
First, we identify topological boundary states as non-Hermitian systems with point-gap topology~\cite{JYLee-19, Bessho-21}.
The Altland-Zirnbauer symmetry for original Hermitian systems reduces to the Altland-Zirnbauer$^{\dag}$ symmetry for non-Hermitian systems~\cite{KSUS-19}, based on chiral symmetry in Eq.~(\ref{eq: CS}), as well as time-reversal symmetry$^{\dag}$ and particle-hole symmetry$^{\dag}$,
\begin{align}
    \mathcal{T} H^{T} \left( \bm{k} \right) \mathcal{T}^{-1} &= H \left( - \bm{k} \right),\\
    \mathcal{C} H^{*} \left( \bm{k} \right) \mathcal{C}^{-1} &= - H \left( -\bm{k} \right),
\end{align}
with unitary matrices $\mathcal{T}$ and $\mathcal{C}$.
This identification is generally feasible as long as point-gap topology is preserved.

Then, the detachability of the topological boundary states reduces to whether the corresponding point-gap topology is intrinsic.
This is classified based on homomorphisms from imaginary-line-gap topology to point-gap topology~\cite{OKSS-20, Shiozaki, imaginary, supplement}.
According to this classification, all the topological boundary states in the Wigner-Dyson classes (i.e., classes A, AI, and AII) cannot be detached from the bulk bands, consistent with their spectral flow~\cite{Nomura-07}. 
By contrast, in the chiral or Bogoliubov-de Gennes classes, some topological boundary states can be detached. 
This feature arises from extrinsic point-gap topology, further reducing to trivial phases in the Wigner-Dyson classes by ignoring chiral and particle-hole symmetries.
Moreover, as a direct consequence of the bulk topology, the detached boundary states exhibit imaginary-line-gap topology, as summarized in Table~\ref{tab: topo inv}.
Similarly to the three-dimensional bulk in class AIII, the boundary topology is captured by $H_{\rm bdy} \left( \bm{k} \right)$, for which particle-hole symmetry behaves like time-reversal symmetry [i.e., $\mathcal{C} H_{\rm bdy}^{*} \left( \bm{k} \right) \mathcal{C}^{-1} = H_{\rm bdy} \left( -\bm{k} \right)$].
Such a projected Hamiltonian can be introduced even for purely Hermitian detached boundary states~\cite{infinite, supplement}. 
In contrast to Ref.~\cite{Altland-24}, our formulation directly captures the detachability of topological boundary states and further facilitates the classification of detached boundary states.

Our classification also predicts other possible topological phases that host detachable boundary states.
For example, surface states in three-dimensional $\mathbb{Z}_2$ topological insulators for class CII are also detachable from the bulk bands.
As an advantage of our formalism, topology of the detached boundary states is systematically identified based on the classification with respect to imaginary line gaps.
In the projected subspace of the detached surface states in class CII, particle-hole symmetry in the original space behaves like time-reversal symmetry, leading to the $\mathbb{Z}_2$ Kane-Mele or Fu-Kane topological invariant (see also Table~\ref{tab: topo inv})~\cite{Kane-Mele-05a, *Kane-Mele-05b, Fu-Kane-06, QHZ-08, Teo=Kane-10, CTSR-review}.
Similarly, in four dimensions, topological boundary states in the Wigner-Dyson classes cannot be separated from the bulk, while those in the chiral and Bogoliubov-de Gennes classes (i.e., classes CII and C) are detachable.
Additionally, in classes DIII and CI, since particle-hole transformation flips chirality, the bulk winding number $W_{2n+1} \left[ \mathcal{H}_{\rm bulk} \right]$ should be even if the boundary states are detached, resulting in the ``$\mathbb{Z}^{\checkmark/\times}$ classification" in Table~\ref{tab: classification}.

\renewcommand{\arraystretch}{1.3}
\newcommand{\bdy}{bdy}

\begin{table}[t]
	\centering
	\caption{Correspondence of topological invariants (top. invs.) between the $d$-dimensional bulk and $\left( d-1 \right)$-dimensional detached boundary (bdy).  
    The symmetry classes are specified for the bulk.
    $\mathbb{Z}$ invariants $W$ for the symmetry classes accompanying chiral symmetry (i.e., classes AIII, BDI, DIII, CII, and CI), and $\mathbb{Z}_2$ invariants $\nu$ for the chiral and Bogoliubov-de Gennes classes.}
	\label{tab: topo inv}
     \begin{tabular}{cc} \hline \hline
        Class & $\left( d-1 \right)$-dim. boundary top. inv. $=$ $d$-dim. bulk top. inv. \\ \hline 
        Chiral & ${\rm Ch}_n\,[H^{(+)}_{\rm \bdy} ]- {\rm Ch}_n\,[H^{(-)}_{\rm \bdy}] = W_{2n+1}\left[\mathcal{H}_{\rm bulk} \right]$ \\ \hline
        BDI & $\nu^{d-1=8n+6}_{\rm AI}\,[H_{\rm \bdy}] = \nu^{d=8n+7}_{\rm BDI}\left[\mathcal{H}_{\rm bulk}\right]$ \\
        BDI \& D & $\nu^{d-1=8n+7}_{\rm AI}\,[H_{\rm \bdy}] = \nu^{d=8n+8}_{\rm BDI \& D}\left[\mathcal{H}_{\rm bulk}\right]$ \\ 
        D & ${\rm Ch}_{4n}\,[H_{\rm \bdy}] \equiv \nu^{d=8n+1}_{\rm D}\left[\mathcal{H}_{\rm bulk}\right]~~(\mathrm{mod}~2)$ \\
        DIII & $\frac{1}{2}{\rm Ch}_{4n}\,[H_{\rm \bdy}] \equiv \nu^{d=8n+1}_{\rm DIII}\left[\mathcal{H}_{\rm bulk}\right]~~(\mathrm{mod}~2)$ \\
        CII & $\nu^{d-1=8n+2}_{\rm AII}\,[H_{\rm \bdy}] = \nu^{d=8n+3}_{\rm CII}\left[\mathcal{H}_{\rm bulk}\right]$ \\
        C \& CII & $\nu^{d-1=8n+3}_{\rm AII}\,[H_{\rm \bdy}] = \nu^{d=8n+4}_{\rm C \& CII}\left[\mathcal{H}_{\rm bulk}\right]$ \\
        C & ${\rm Ch}_{4n+2}\,[H_{\rm \bdy}] \equiv \nu^{d=8n+5}_{\rm C}\left[\mathcal{H}_{\rm bulk}\right]~~(\mathrm{mod}~2)$ \\
        CI & $\frac{1}{2}{\rm Ch}_{4n+2}\,[H_{\rm \bdy}] \equiv \nu^{d=8n+5}_{\rm CI}\left[\mathcal{H}_{\rm bulk}\right]~~(\mathrm{mod}~2)$ \\ \hline \hline
  \end{tabular}
\end{table}

\renewcommand{\arraystretch}{1}

\textit{Wannier localizability}.---Whereas we have focused on the 
detachability of topological boundary states,
we now clarify its connection with the bulk Wannier localizability, using $K$-theory~\cite{Karoubi} (see Fig.~\ref{fig: commutative diagram}).
For a gapped $d$-dimensional Hermitian system, we consider the bulk Hamiltonian $\mathcal{H}_{\rm bulk}\,(\bk)$ with periodic boundary conditions and the boundary Hamiltonian $\hat{\mathcal{H}}_{\rm bdy}\,(\bk_\parallel)$ with semi-infinite boundary conditions~\cite{infinite}. 
The corresponding $K$-groups underlying their topological classifications are related by the bulk-boundary correspondence~\cite{Karoubi, AS-69, FM-13, Thiang-16, Shiozaki-17, Gomi-21}.
As explained previously,
the boundary Hamiltonian $\hat{\mathcal{H}}_{\rm bdy}\,(\bk_\parallel)$ exhibits no spectral flow and is thus detachable if and only if it is in trivial phases within the corresponding Wigner-Dyson classes by ignoring chiral and particle-hole symmetries.
From a bulk perspective, by contrast, this reduction to the Wigner-Dyson classes dictates the presence or absence of exponentially localized bulk Wannier functions. 
In fact, if $\hat{\mathcal{H}}_{\rm bdy}\,(\bk_\parallel)$ exhibits spectral flow on the boundary, $\mathcal{H}_{\rm bulk}\left(\bk\right)$ holds a nontrivial topological invariant even when chiral and particle-hole symmetries are ignored, which obstructs the exponentially localized Wannier functions.
Conversely, the absence of boundary spectral flow permits exponential Wannier localization \cite{Altland-24}.
We further elaborate on the $K$-theory classification in our accompanying work~\cite{SNSSK-24}.

\textit{Discussions}.---In this Letter, we elucidate that non-Hermitian topology underlies Wannier localizability and detachable boundary states in Hermitian topological materials, thereby establishing their classification (Table~\ref{tab: classification}).
We demonstrate that while intrinsic non-Hermitian topology prohibits the detachment of topological boundary states, extrinsic non-Hermitian topology permits it and further leads to anti-Hermitian topology of the detached boundary states.
Our formulation reveals that the detached topological boundary states concomitantly exhibit Wigner-Dyson-type gapped topology in total and chiral-type or Bogoliubov-de Gennes-type gapless topology at zero energy.
They should exhibit unique Anderson localization and transitions, meriting further investigation.
Our framework may also be useful in engineering topological flat bands at boundaries.
The precise conditions to realize boundary topological flat bands are left for future work.

Our non-Hermitian perspective directly characterizes the detachability of topological boundary states and complements the bulk classification of Wannier localizability~\cite{Altland-24}.
This complementary viewpoint should be relevant to extending our results to topological crystalline insulators, where the bulk-boundary correspondence becomes more subtle.
As a prime example, we investigate reflection-invariant second-order topological insulators~\cite{Chiu-13, Morimoto-13, Shiozaki-14, Langbehn-17, supplement}.
It is also worthwhile to generalize our formulation to detachable boundary states in other types of topological insulators, such as Hopf insulators~\cite{Moore-08, Alexandradinata-21} and nonsymmorphic insulators~\cite{Fang-15, Shiozaki-15, *Shiozaki-16}.
Finally, a possible connection with topological unitary operators~\cite{Gong-21} is worth noting.

\medskip
{\it Note added}.---After the completion of this work, we became aware of a recent related work~\cite{Lapierre-24}.

\medskip
\begingroup
\renewcommand{\addcontentsline}[3]{}
\begin{acknowledgments}
We thank Akira Furusaki, Zongping Gong, Shu Hamanaka, Shingo Kobayashi, and Shinsei Ryu for helpful discussion.
We appreciate the long-term workshops ``Recent Developments and Challenges in Topological Phases" (YITP-T-24-03) and ``Dynamics Days Asia Pacific 13 (DDAP13) / YKIS2024" held at Yukawa Institute for Theoretical Physics (YITP), Kyoto University.
D.N., K. Shiozaki, K. Shimomura, and M.S. are supported by JST CREST Grant No.~JPMJCR19T2. 
D.N. is supported by JSPS KAKENHI Grant No.~JP24K22857.
K. Shiozaki is supported by JSPS KAKENHI Grants No.~JP22H05118 and No.~JP23H01097. 
K. Shimomura is supported by JST SPRING, Grant No.~JPMJSP2110 and JSPS KAKENHI Grant No.~JP25KJ1632.
M.S. is supported by JSPS KAKENHI Grants No.~JP24K00569 and No.~JP25H01250. 
K.K. is supported by MEXT KAKENHI Grant-in-Aid for Transformative Research Areas A ``Extreme Universe" No.~JP24H00945.
\end{acknowledgments}
\endgroup

\let\oldaddcontentsline\addcontentsline
\renewcommand{\addcontentsline}[3]{}
\bibliography{NH_top.bib}

\begin{thebibliography}{117}
\bibitem [{pH()}]{pH}%
  \BibitemOpen
  \href@noop {} {}\bibinfo {note} {{This $\mathbb{Z}_2$ subsector structure due to chiral symmetry (or equivalently, pseudo-Hermiticity) is clarified by the combination of right and left eigenstates; see Sec.~VI\,B in Ref.~\cite{KSUS-19}.}}\BibitemShut {Stop}%
\bibitem [{ima()}]{imaginary}%
  \BibitemOpen
  \href@noop {} {}\bibinfo {note} {{In the Altland-Zirnbauer$^{\dag}$ symmetry classes, all extrinsic point-gap topology reduces to imaginary-line-gap topology. If it were instead reducible to real-line-gap topology, the associated topological boundary states could open a gap even around zero energy, thereby contradicting their stability ensured by the bulk Hermitian topology.}}\BibitemShut {Stop}%
\bibitem [{inf()}]{infinite}%
  \BibitemOpen
  \href@noop {} {}\bibinfo {note} {{Even if the boundary Hamiltonians are infinite dimensional, the boundary projectors and concomitant topological invariants can be formulated with finite-dimensional Bloch states.}}\BibitemShut {Stop}%
\end{thebibliography}
\let\addcontentsline\oldaddcontentsline

\clearpage
\widetext

\setcounter{secnumdepth}{3}

\renewcommand{\theequation}{S\arabic{equation}}
\renewcommand{\thefigure}{S\arabic{figure}}
\renewcommand{\thetable}{S\arabic{table}}
\setcounter{equation}{0}
\setcounter{figure}{0}
\setcounter{table}{0}
\setcounter{section}{0}
\setcounter{tocdepth}{0}

\numberwithin{equation}{section} 

\begin{center}
{\bf \large Supplemental Material for \\ \smallskip 
``Non-Hermitian Origin of Detachable Boundary States in Topological Insulators"
}
\end{center}



\section{Detachment of topological boundary states}
    \label{asec: detachment}

\subsection{2D class A}
    \label{asubsec: 2D class A}

We consider a one-dimensional Dirac model
\begin{equation}
    \mathcal{H} \left( k_x \right) = k_x,
\end{equation}
for a chiral edge state of a two-dimensional topological insulator without symmetry (i.e., class A).
An important feature of this Dirac model is spectral flow.
To see this, we couple it with a trivial band $\mathcal{H} = 0$,
\begin{equation}
    \mathcal{H} \left( k_x \right) = \begin{pmatrix}
        k_x & v \\
        v^{*} & 0
    \end{pmatrix},
\end{equation}
with the coupling amplitude $v \in \mathbb{C}$.
The spectrum of this two-band model is obtained as
\begin{equation}
    E \left( k_x \right) = \frac{k_x}{2} \pm \sqrt{\left( \frac{k_x}{2} \right)^2 + \left| v \right|^2},
\end{equation}
which behaves
\begin{equation}
    E \left( k_x \right) \rightarrow k_x, - \frac{\left| v \right|^2}{k_x} \quad \left( k \to \pm \infty \right).
\end{equation}
Thus, such a perturbation cannot open a spectral gap, implying the inevitable presence of the spectral flow.

\subsection{2D class AII}

We consider a one-dimensional Dirac model
\begin{equation}
    \mathcal{H} \left( k_x \right) = k_x \sigma_x, 
        \label{aeq: 2D class AII}
\end{equation}
which respects time-reversal symmetry
\begin{equation}
    \sigma_y \mathcal{H}^{*} \left( k_x \right) \sigma_y^{-1} = \mathcal{H} \left( -k_x \right).   
        \label{aeq: 2D class AII TRS}
\end{equation}
Similarly to class A, this Dirac model exhibits spectral flow.
To see this, we again couple it with a trivial band $\mathcal{H}=\varepsilon \in \mathbb{R}$ so that time-reversal symmetry will be respected,
\begin{equation}
    \mathcal{H} \left( k_x \right) = \begin{pmatrix}
        k_x \sigma_x & -\ii v \sigma_x \\
        \ii v \sigma_x & \varepsilon
    \end{pmatrix} 
    = \frac{1}{2} \left( k_x \sigma_x + \varepsilon \right) + \frac{1}{2} \left( k_x \sigma_x - \varepsilon \right) \tau_z + v \sigma_x \tau_y,
\end{equation}
with the coupling amplitude $v \in \mathbb{C}$.
Since this four-band model commutes with $\sigma_x$, it consists of two independent two-band models in class A,
\begin{equation}
    \begin{pmatrix}
        k_x & -\ii v \\
        \ii v & \varepsilon
    \end{pmatrix},\quad \begin{pmatrix}
        -k_x & \ii v \\
        -\ii v & \varepsilon
    \end{pmatrix},
\end{equation}
each of which exhibits spectral flow, as discussed in Sec.~\ref{asubsec: 2D class A}.

The unitary symmetry that commutes with the Hamiltonian is specific to the above model and not necessarily present in generic models.
Nevertheless, the Dirac model generally accompanies spectral flow as long as time-reversal symmetry in Eq.~(\ref{aeq: 2D class AII TRS}) is respected~\cite{Nomura-07}.
Let us put this Dirac model on a one-dimensional space with length $L_x$ and under the twisted boundary conditions.
For the twist angle $\phi$, momentum $k_x$ is quantized by
\begin{equation}
    e^{\ii k_x L_x} = e^{\ii \phi}, \quad \mathrm{i.e.}, \quad k_x L_x = \phi + 2n\pi \quad \left( n \in \mathbb{Z} \right),
\end{equation}
and hence the energy spectrum of the Dirac model in Eq.~(\ref{aeq: 2D class AII}) is obtained as
\begin{equation}
    E \left( \phi \right) = \pm \frac{\phi + 2n\pi}{L_x}.
\end{equation}
An important consequence of time-reversal symmetry in Eq.~(\ref{aeq: 2D class AII TRS}) is the Kramers degeneracy for $\phi \in \pi\mathbb{Z}$.
On the other hand, for $\phi \notin \pi\mathbb{Z}$, time-reversal symmetry is broken, resulting in no Kramers degeneracy.
Through the adiabatic change of the flux $\phi$, each of the original Kramers pairs for $\phi = 0$ splits for $0 < \phi < \pi$ and then meets the other half of the other Kramers pairs.
Such switches of the Kramers pairs inevitably require the energy cutoff in high energy and the concomitant spectral flow.
It is notable that this discussion is also applicable to two-dimensional and three-dimensional Dirac models with time-reversal symmetry in Eq.~(\ref{aeq: 2D class AII TRS}).
Additionally, it is applicable even in the presence of disorder that breaks translation invariance.

\subsection{3D class AIII}
    \label{asubsec: boundary 3D class AIII}

We consider a two-dimensional Dirac model
\begin{equation}
    \mathcal{H} \left( k_x, k_y \right) = k_x \sigma_x + k_y \sigma_y,
        \label{aeq: 2D AIII}
\end{equation}
which respects chiral symmetry
\begin{equation}
    \sigma_z \mathcal{H} \left( k_x, k_y \right) \sigma_z^{-1} = - \mathcal{H} \left( k_x, k_y \right).
        \label{aeq: 2D AIII chiral}
\end{equation}
This continuum model describes a surface state of a three-dimensional topological insulator in class AIII.
Notably, this chiral-symmetric Dirac model exhibits no spectral flow~\cite{Altland-24}.
To confirm this, we couple it to a trivial band and study the four-band model,
\begin{align}
    \mathcal{H} \left( k_x, k_y \right) &= \begin{pmatrix}
        k_x \sigma_x + k_y \sigma_y & v\sigma_-  \\
        v \sigma_+ & \varepsilon \sigma_x
    \end{pmatrix} \nonumber \\
    &= \frac{1}{2} \left( \left( k_x + \varepsilon \right) \sigma_x + k_y \sigma_y \right) + \frac{1}{2} \left( \left( k_x - \varepsilon \right) \sigma_x + k_y \sigma_y \right) \tau_z + \frac{v}{2} \left( \sigma_x \tau_x + \sigma_y \tau_y \right),
\end{align}
where $\varepsilon \sigma_x$ ($\varepsilon \geq 0$) is a trivial band, and $v \geq 0$ denotes the coupling strength to the Dirac model.
The entire four-band model remains to respect chiral symmetry in Eq.~(\ref{aeq: 2D AIII chiral}).
Since we have 
\begin{align}
    2\mathcal{H}^2 \left( k_x, k_y \right) = k_x^2 + k_y^2 + \varepsilon^2 + v^2 + v \left(  k_x + \varepsilon + \left( k_x - \varepsilon \right) \sigma_z \right) \tau_x + v k_y \left( 1 + \sigma_z \right) \tau_y + \left( k_x^2 + k_y^2 - \varepsilon^2 -v^2 \sigma_z \right) \tau_z,
\end{align}
$H^2$ commutes with $\sigma_z$.
\begin{itemize}
    \item For $\sigma_z = +1$, we have
    \begin{equation}
        2\mathcal{H}^2 \left( k_x, k_y \right) = k_x^2 + k_y^2 + \varepsilon^2 + v^2 + 2v k_x \tau_x + 2v k_y \tau_y + \left( k_x^2 + k_y^2 - \varepsilon^2 -v^2 \right) \tau_z,
    \end{equation}
    and hence the energy bands are determined by 
    \begin{align}
        2E^2 \left( k_x, k_y \right) &= k^2 + \varepsilon^2 + v^2 \pm \sqrt{\left( k^2 - \varepsilon^2- v^2 \right)^2 + 4v^2 k^2} \nonumber \\
        &= k^2 + \varepsilon^2 + v^2 \pm \sqrt{\left( k^2 - \varepsilon^2 + v^2 \right)^2 + 4\varepsilon^2 v^2}
            \label{aeq: 2D AIII E^2}
    \end{align}
    with $k^2 \coloneqq k_x^2 + k_y^2$.

    \item For $\sigma_z = -1$, we have
    \begin{equation}
        2\mathcal{H}^2 \left( k_x, k_y \right) = k_x^2 + k_y^2 + \varepsilon^2 + v^2 + 2 \varepsilon v \tau_x + \left( k_x^2 + k_y^2 - \varepsilon^2 +v^2 \right) \tau_z,
    \end{equation}
    and hence the energy bands are again determined by Eq.~(\ref{aeq: 2D AIII E^2}).
\end{itemize}
Therefore, regardless of $\sigma_z$, the four energy bands are obtained as
\begin{equation}
    \sqrt{2} E \left( k \right) = \pm \sqrt{ k^2 + \varepsilon^2 + v^2 \pm \sqrt{\left( k^2 - \varepsilon^2+ v^2 \right)^2 + 4\varepsilon^2v^2} }.
\end{equation}
Around the Dirac point (i.e., $\left| k_x \right|, \left| k_y \right| \ll \varepsilon, v$), we have
\begin{align}
    E \left( k \right) = \pm \sqrt{\varepsilon^2 + v^2} + O \left( k^2 \right), \pm \frac{\varepsilon}{\sqrt{\varepsilon^2 + v^2}} k + O \left( k^3 \right).
\end{align}
In the ultraviolet limit (i.e., $\left| k_x \right|, \left| k_y \right| \to \infty$), on the other hand, we have
\begin{align}
    E \left( k \right) = \pm \sqrt{k^2 + v^2 + O \left( 1/k^2 \right)}, \pm \varepsilon + O \left( 1/k^2 \right).
\end{align}
Thus, the four-band model exhibits a global gap
\begin{equation}
    \Delta = \sqrt{\varepsilon^2 + v^2} - \varepsilon.
        \label{aeq: 3D AIII - global gap}
\end{equation}

\section{Point-gap topology and exceptional points}

\begin{figure}[H]
 \centering
  \includegraphics[width = 0.8\linewidth]{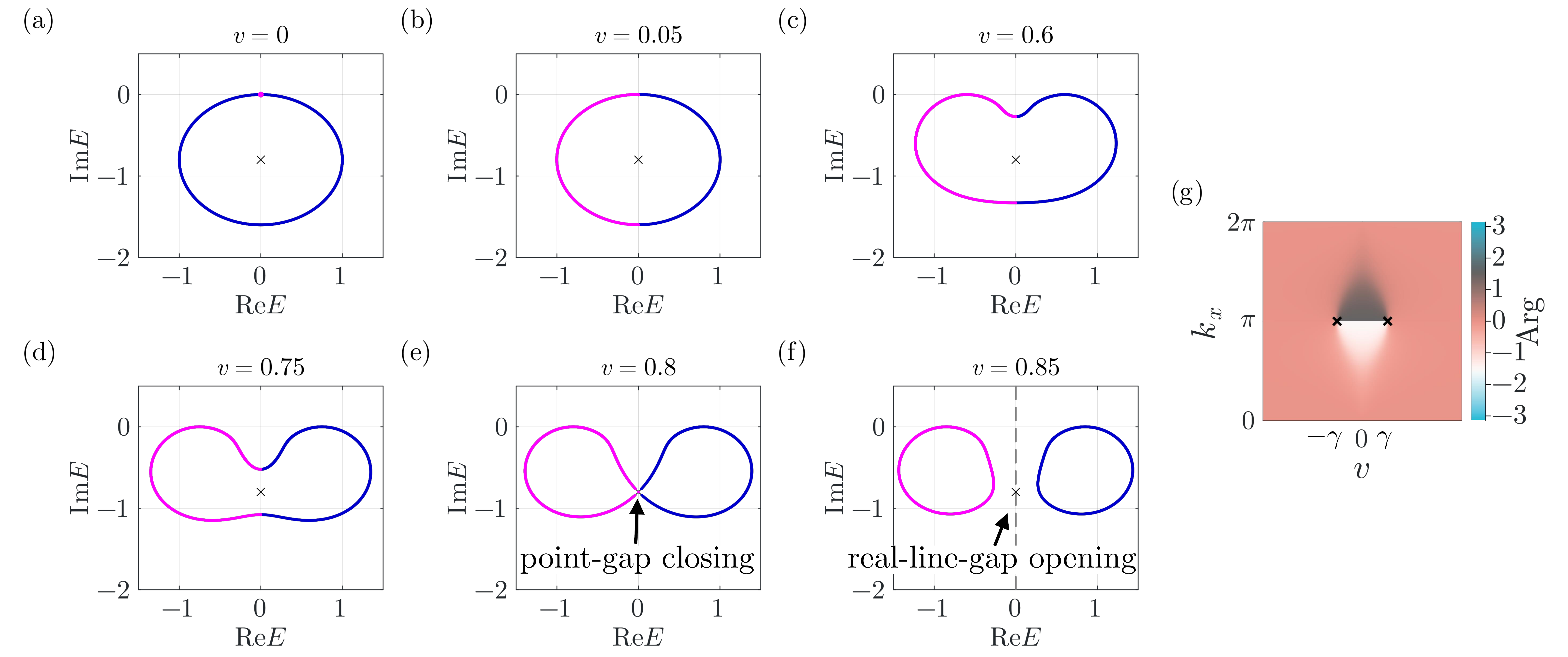}
    \caption{Complex spectrum of the non-Hermitian model in Eq.~(\ref{aeq: NH chiral}) for $\gamma = 0.8$ and (a)~$v=0$, (b)~$v=0.05$, (c)~$v=0.6$, (d)~$v=0.75$, (e)~$v=0.8$, and (f)~$v=0.85$.
    The two bands are denoted by blue and magenta.
    (g)~Argument of $E_{+}-E_{-}$ as a function of $v$ and $k_x$, where $E_{\pm} \coloneqq (H_{\rm A}  \pm \sqrt{H_{\rm A}^2 + 4\,| v |^2}
    )/2$ represent the two complex eigenenergies of $\tilde{H}_{\rm A}$.
    A pair of exceptional points (black crosses) appears at $\left( v, k_x \right) = \left( \pm \gamma, \pi \right)$, where the argument is ill defined.
    These exceptional points are connected through the branch cut, which appears even for $v=0$ owing to the definitions of $E_{\pm}$, involving the square root of the complex function, while the entire spectrum is consistent with $\tilde{H}_{\rm A}$.}
    \label{afig: NH chiral}
\end{figure}

In Fig.~\ref{afig: NH chiral}, we show the complex spectrum of the non-Hermitian model [i.e., Eq.~(\ref{eq: 1D class A perturbation}) in the main text]
\begin{equation}   
    \tilde{H}_{\rm A} \left( k_x \right) = \begin{pmatrix}
        H_{\rm A} \left( k_x \right) & v \\
        v^{*} & 0
    \end{pmatrix}, \quad H_{\rm A} \left( k_x \right) = \sin k_x + \ii \gamma \left( \cos k_x - 1 \right),
        \label{aeq: NH chiral}
\end{equation}
which describes a non-Hermitian counterpart of the chiral edge state with an additional trivial band.
For the sufficiently weak coupling strength $\left| v \right| < \gamma$, a point gap around $E = -\ii \gamma$ remains open, accompanied by the nontrivial point-gap topology $W_1 = 1$ (i.e., winding number of the complex spectrum)~\cite{Gong-18, KSUS-19}.
Since this point-gap topology is intrinsic to non-Hermitian systems and irreducible to line-gap topology, 
it necessitates the swapping of the two bands as momentum $k_x$ is traversed through the Brillouin zone [see Fig.~\ref{afig: NH chiral}\,(b-d)].
Meanwhile, the point gap closes at $\left| v \right| = \gamma$, making the point-gap topology around $E = -\ii \gamma$ trivial for $\left| v \right| > \gamma$ [see Fig.~\ref{afig: NH chiral}\,(e, f)].
In such a case, this non-Hermitian system can no longer be identified as a chiral edge state.

The swapping of complex energy bands is closely related to exceptional points [Fig.~\ref{afig: NH chiral}\,(g)].
In the parameter space $\left( v, k_x \right)$, a pair of exceptional points appears at $\left( v, k_x \right) = \left( \pm \gamma, \pi \right)$, connected by a branch cut.
For $\left| v \right| < \gamma$, traversing the Brillouin zone in momentum space entails crossing this branch cut, resulting in the exchange of the two bands.

\section{From point gap to imaginary line gap}
    \label{asec: 2D AIII PL}

We provide a one-parameter point-gapped model in two dimensions for class AIII that shows imaginary-line-gap opening while retaining a point gap: 
\begin{align}
    H_\theta \left( k_x, k_y \right) &\coloneqq \Gamma U(\theta) \Gamma h\left( k_x, k_y \right) U^{\dagger}(\theta) - \ii \sigma_0\tau_0 \quad \left( \Gamma \coloneqq \sigma_z \tau_0\right),
        \label{aeq: 2D AIII cont. defo.}
\end{align}
with a one parameter $\theta\in[0,1]$ and 
\begin{align}
    U(\theta) &\coloneqq \textrm{exp}\left[\dfrac{\ii\pi\theta}{4}\sigma_y(\tau_0+\tau_z)\right]\textrm{exp}\left[\dfrac{\ii\pi\theta}{4}(\sigma_0+\sigma_z)\tau_y\right],\\
    h\left( k_x, k_y \right) &\coloneqq \dfrac{1}{\mathcal{N}}\left[\sin k_x \sigma_x + \left(1 - \cos k_x - \cos k_y\right) \sigma_y + \ii\sin k_y \sigma_0\right]\dfrac{\tau_0 + \tau_z}{2} + \ii\sigma_0\dfrac{\tau_0 - \tau_z}{2},\\
    \mathcal{N} &\coloneqq \sqrt{\sin^2 k_x + \left(1 - \cos k_x - \cos k_y\right)^2 + \sin^2 k_y}.
\end{align}
For arbitrary $\theta \in \left[ 0, 1 \right]$, this model indeed respects chiral symmetry:
\begin{equation}
    \Gamma H_\theta^{\dagger} \Gamma^{-1} = - H_\theta.
\end{equation}

For $\theta=0$, this model reduces to 
\begin{equation}
    H_{0} \left( k_x, k_y \right) = \dfrac{1}{\mathcal{N}}\left[\sin k_x \sigma_x + \left(1 - \cos k_x - \cos k_y\right) \sigma_y + \ii (\sin k_y-\mathcal{N}) \sigma_0\right]\dfrac{\tau_0 + \tau_z}{2},
\end{equation}
which is nothing but Eq.~(\ref{eq: 2dAIII}) in the main text with $\gamma=1$, up to normalization and stacking with trivial bands. 
$H_0$ has a point gap around $E_{\mathrm{P}}=-\ii$ and exhibits nontrivial point-gap topology (i.e., $\mathrm{Ch}_1 \left[\ii \left( H_{0}-E_{\mathrm{P}} \right) \Gamma \right] \neq 0$), as explained in the main text. 
Since $\Gamma$ and $U(\theta)$ are unitary matrices, it is straightforward to confirm that $H_\theta$ keeps the point gap at $E_{\mathrm{P}}$ and chiral symmetry even for arbitrary $0<\theta\leq1$.

\begin{figure}[H]
 \begin{center}
 \hspace*{-3em}
  \includegraphics[scale=0.13]{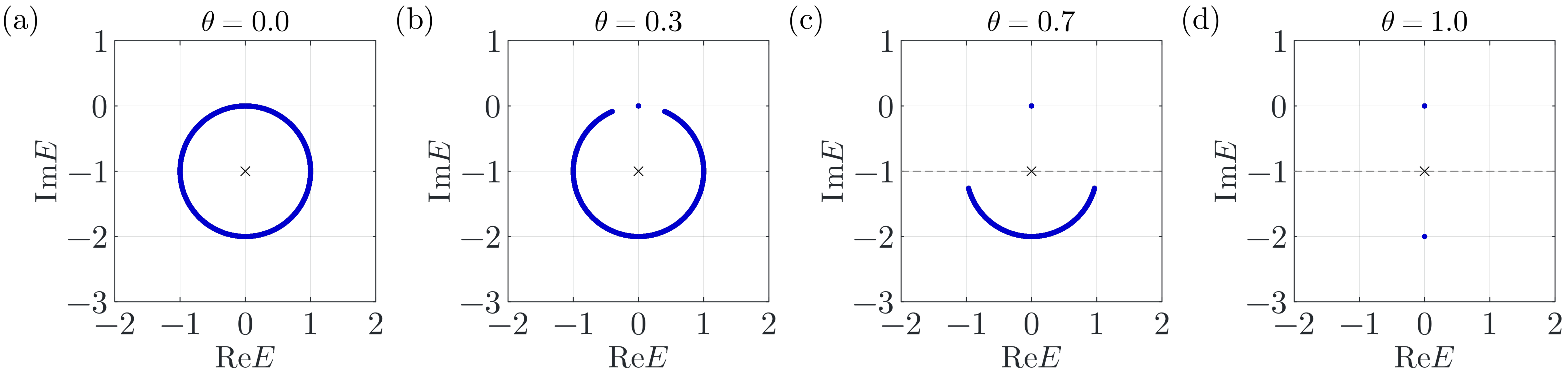}
  \end{center}
    \caption{Opening of an imaginary line gap while keeping a point gap.
    The complex spectrum of the point-gapped two-dimensional model $H_{\theta} \left( k_x, k_y \right)$ in class AIII [Eq.~(\ref{aeq: 2D AIII cont. defo.})] is shown for (a)~$\theta = 0.0$, (b)~$\theta = 0.3$, (c)~$\theta = 0.7$, and (d)~$\theta = 1.0$.}
    \label{fig: 2D class AIII cont. defo.}
\end{figure}

As illustrated in Fig.~\ref{fig: 2D class AIII cont. defo.}, with increasing $\theta$, eigenstates with $\mathrm{Im}\,E<0$ leave from those with $E=0$, leading to opening of an imaginary line gap with respect to $E_{\mathrm{L}}=-1$. 
Finally, $H_{\theta}$ reaches an anti-Hermitian Chern insulator at $\theta=1$ while keeping the point gap at $E_{\mathrm{P}}$:
\begin{equation}
    H_1 \left( k_x, k_y \right) = \dfrac{\sigma_0 + \sigma_z}{2}\dfrac{\ii}{\mathcal{N}}\left[\sin k_x \tau_y + \left(1 - \cos k_x - \cos k_y\right) \tau_x - \sin k_y \tau_z\right] - \dfrac{\sigma_0 - \sigma_z}{2}\ii\tau_z - \ii\sigma_0\tau_0.
\end{equation}

\begin{figure}[H]
\centering
\includegraphics[width=0.7\linewidth]{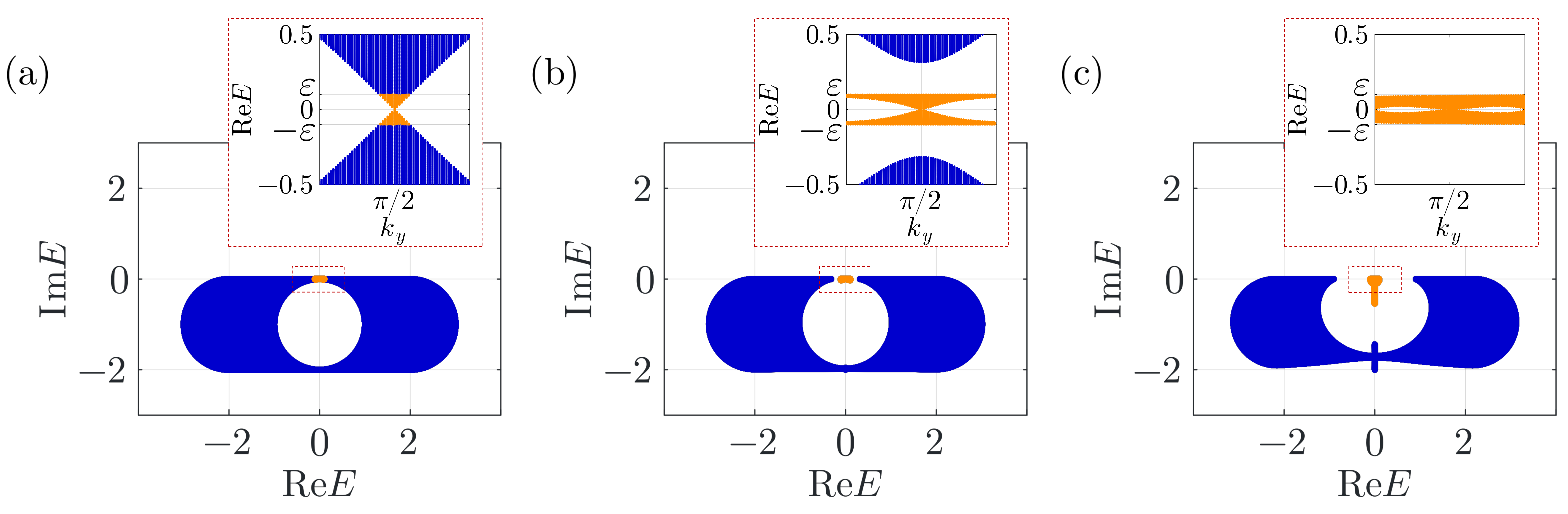}
\caption{Complex spectrum of the non-Hermitian model in Eq.~(\ref{aeq: 2D class AIII - main text}) for $\gamma = 1$, $\varepsilon = 0.1$, as well as (a)~$v = 0.0$, (b)~$v = 0.3$, and (c)~$v = 0.9$.
The Dirac surface state around $E=0$ (orange) can be detached from the bulk bands (blue) in the presence of the perturbation $\varepsilon\sigma_x$.}
    \label{afig: 2D class AIII cont. defo. - 2}
\end{figure}

In Fig.~\ref{afig: 2D class AIII cont. defo. - 2}, we also show the complex spectrum of the non-Hermitian model [i.e., Eq.~(\ref{eq: 2D class AIII perturbation}) in the main text]
\begin{equation}
    \tilde{H}_{\rm AIII} \left( \bm{k} \right) = \begin{pmatrix}
        H_{\rm AIII} \left( \bm{k} \right) & v \sigma_- \\
        v^{*}\sigma_+ & \varepsilon \sigma_x
    \end{pmatrix}, \quad H_{\rm AIII} \left( \bm{k} \right) = \left( \sin k_x \right) \sigma_x + \left( 1- \cos k_x -\cos k_y \right) \sigma_y + \ii \gamma \left( \sin k_y - 1 \right),
        \label{aeq: 2D class AIII - main text}
\end{equation}
which also shows the continuous deformation from point-gap topology to imaginary-line-gap topology.

\section{Bulk-boundary correspondence}
    \label{asec: BBC}

\begin{figure}[H]
    \centering
    \includegraphics[scale=0.04]{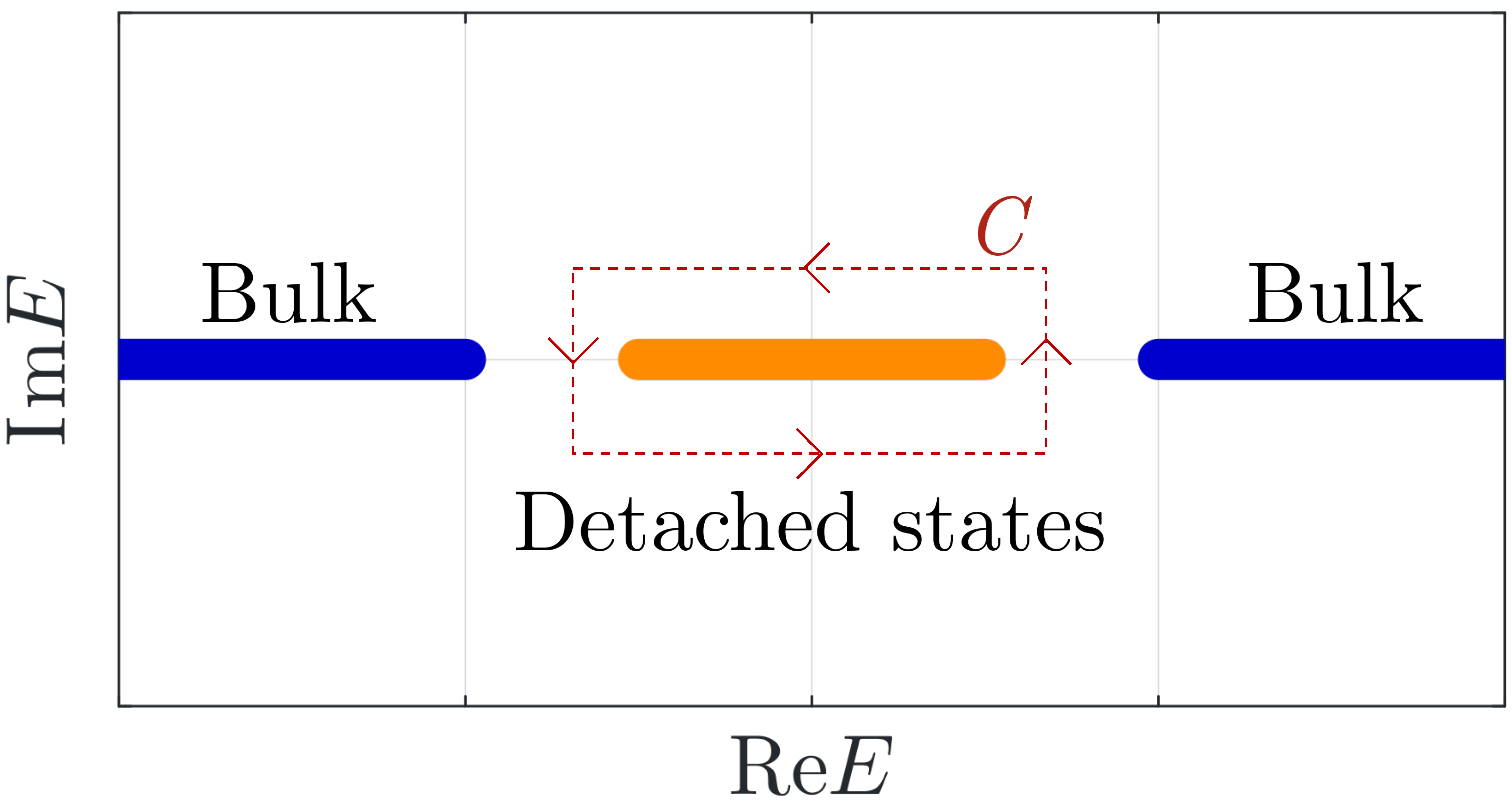}
    \caption{Projection onto the detached boundary states. 
    The loop $C$ is chosen to respect symmetry and enclose the spectrum of the detached boundary states in the complex energy plane.}
        \label{fig: projection}
\end{figure}

Our classification uncovers the topological nature of the detached boundary states. 
For a gapped $d$-dimensional Hermitian Hamiltonian $\mathcal{H}_{\rm bulk}\,(\bk)$, the bulk topological invariant is defined~\cite{Schnyder-08,CTSR-review}. 
For the semi-infinite boundary conditions with a boundary perpendicular to the $x_d$ direction, the boundary Hamiltonian $\hat{\mathcal{H}}_{\rm bdy}\,(\bk_\parallel)$ is defined, where $\bk_{\parallel} = (k_1,\dots,k_{d-1})$ is the Bloch momentum for the surface Brillouin zone. 
Note that the matrix size of $\hat{\mathcal{H}}_{\rm bdy}\,(\bk_\parallel)$ is infinite due to the bulk degrees of freedom. 
Suppose that the boundary Hamiltonian $\hat{\mathcal{H}}_{\rm bdy}\,(\bk_\parallel)$ hosts states detached from the bulk spectrum; 
the orthogonal projection onto the detached states is defined as 
\begin{align}
    \hat P\,(\bk_\parallel) = \frac{1}{2\pi \ii} \oint_{C} \frac{dz}{z-\hat{\mathcal{H}}_{\rm bdy}\,(\bk_\parallel)},
\end{align}
where $C$ is a symmetry-preserving loop in the complex energy plane surrounding the spectrum of the detached states (see Fig.~\ref{fig: projection}). 
$\hat P\,(\bk_\parallel)$ sends the bulk state to 0 and the detached boundary states to 1.
Notably, particle-hole and chiral symmetries behave as time-reversal and $\mathbb{Z}_2$ unitary symmetries, respectively, within the sub-Hilbert space of the detached boundary states: 
\begin{align}
    \hat {\cal C}\hat P^*\,(\bk_\parallel)\,\hat {\cal C}^{-1} &= \hat P\,(-\bk_\parallel) \quad (\hat {\cal C}\hat {\cal C}^*=\pm 1), \\
    \hat \Gamma \hat P\,(\bk_\parallel) \,\hat \Gamma^{-1} &= \hat P\,(\bk_\parallel)
\end{align}
with unitary operators $\hat {\cal C}$ and $\hat \Gamma$ for particle-hole and chiral symmetries.
Accordingly, we define the $(d-1)$-dimensional topological invariants for the detached boundary states protected by $\hat {\cal C}$ and $\hat \Gamma$, as specified below.

For odd $d = 2n+1$ in the presence of chiral symmetry, with the projector $\hat P^{(\pm)}\,(\bk_\parallel) \coloneqq \hat P\,(\bk_\parallel) (1 \pm \hat \Gamma)/2$ for each chirality $\hat \Gamma=\pm 1$, two independent Chern numbers are defined as 
\begin{equation}
    {\rm Ch}_n\left[\hat P^{(\pm)}\,(\bk_\parallel)\right] 
    \coloneqq \frac{1}{n!}\left(\frac{\ii}{2\pi}\right)^n \int_{T^{d-1}} {\rm Tr}\left[\left(\hat P^{(\pm)}\,(\bk_\parallel)\,d\hat P^{(\pm)}\,(\bk_\parallel)\,d\hat P^{(\pm)}\,(\bk_\parallel)\right)^n\right].
\end{equation}
In the presence of particle-hole symmetry, as $\hat{\cal C}$ behaves as time-reversal symmetry, one can define $\mathbb{Z}_2$ invariants for the detached boundary states. 
Specifically, one can define the class-AII-type $\mathbb{Z}_2$ invariant $\nu^{d-1}_{\rm class\ AII}\left[\hat P\,(\bk_\parallel)\right]$ for class C and the boundary dimensions $d-1=8n+2, 8n+3$, and the class-AI-type $\mathbb{Z}_2$ invariant $\nu^{d-1}_{\rm class\ AI}\left[\hat P\,(\bk_\parallel)\right]$ for class D and $d-1=8n+6, 8n+7$~\cite{QHZ-08,Ryu-10,Teo=Kane-10}. 
Furthermore, since particle-hole symmetry $\hat {\cal C}$ does not change the chirality (i.e., $\hat {\cal C}\hat \Gamma^* = \hat \Gamma \hat {\cal C}$) for classes CII and BDI, the class-AII-type $\Z_2$ invariant $\nu^{d-1}_{\rm class\ AII}\left[\hat P^{(\pm)}\,(\bk_\parallel)\right]$ for each sector $\hat \Gamma = \pm$ is well defined for class CII and $d-1 = 8n+2, 8n+3$. Similarly, the class-AI-type $\Z_2$ invariant $\nu^{d-1}_{\rm class\ AI}\left[\hat P^{(\pm)}\,(\bk_\parallel)\right]$ for each sector $\hat \Gamma = \pm$ is well defined for class BDI and $d-1 = 8n+6, 8n+7$.

The imaginary-line-gap to point-gap map described in Sec.~\ref{asec: intrinsic} gives the relationship between the bulk and boundary invariants, summarized as follows. 
In the presence of chiral symmetry, we have 
\begin{align}
    {\rm Ch}_n\left[\hat P^{(+)}\,(\bk_\parallel)\right]- {\rm Ch}_n\left[\hat P^{(-)}\,(\bk_\parallel)\right] = W_{2n+1}\left[\mathcal{H}_{\rm bulk}\,(\bk)\right], \label{eq:BBC_chiral}
\end{align}
where $W_{2n+1}\left[\mathcal{H}_{\rm bulk}\,(\bk)\right]$ is the $(2n+1)$-dimensional winding number defined by the chiral-symmetric Hamiltonian $\mathcal{H}_{\rm bulk}\,(\bk)$. 
For classes DIII and CI, since particle-hole symmetry flips chirality as $\hat {\cal C}\hat \Gamma^*=- \hat \Gamma \hat {\cal C}$, we have ${\rm Ch}_n\left[\hat P^{(+)}\,(\bk_\parallel)\right]=-{\rm Ch}_n\left[\hat P^{(-)}\,(\bk_\parallel)\right]$, meaning that if the boundary states are detached, the bulk winding number $W_{2n+1}\left[\mathcal{H}_{\rm bulk}\,(\bk)\right]$ should be even. 
This accounts for the ``$\mathbb{Z}^{\checkmark/\times}$ classification" in Table~\ref{tab: classification}.
For the $\mathbb{Z}_2$ topological phases in the $d$-dimensional bulk, we obtain
\begin{align}
    {\rm Ch}_{4n}\left[ \hat P\,(\bk_\parallel) \right] \equiv \nu^{d=8n+1}_{\rm class\ D}\left[\mathcal{H}_{\rm bulk}\,(\bk)\right] \quad \left( \mathrm{mod}~2\right), \\
    \frac{1}{2}{\rm Ch}_{4n}\left[ \hat P\,(\bk_\parallel) \right] \equiv \nu^{d=8n+1}_{\rm class\ DIII}\left[\mathcal{H}_{\rm bulk}\,(\bk)\right] \quad \left( \mathrm{mod}~2\right), \\
    \nu^{d-1=8n+2}_{\rm class\ AII}\left[\hat P^{(+)}\,(\bk_\parallel)\right]+\nu^{d-1=8n+2}_{\rm class\ AII}\left[\hat P^{(-)}\,(\bk_\parallel)\right] = \nu^{d=8n+3}_{\rm class\ CII}\left[\mathcal{H}_{\rm bulk}\,(\bk)\right], \\
    \nu^{d-1=8n+3}_{\rm class\ AII}\left[\hat P^{(+)}\,(\bk_\parallel)\right]+\nu^{d-1=8n+3}_{\rm class\ AII}\left[\hat P^{(-)}\,(\bk_\parallel)\right] = \nu^{d=8n+4}_{\rm class\ CII}\left[\mathcal{H}_{\rm bulk}\,(\bk)\right], \\
    \nu^{d-1=8n+3}_{\rm class\ AII}\left[\hat P\,(\bk_\parallel)\right] = \nu^{d=8n+4}_{\rm class\ C}\left[\mathcal{H}_{\rm bulk}\,(\bk)\right], \\
    {\rm Ch}_{4n+2}\left[\hat P\,(\bk_\parallel)\right] \equiv \nu^{d=8n+5}_{\rm class\ C}\left[\mathcal{H}_{\rm bulk}\,(\bk)\right] \quad \left( \mathrm{mod}~2\right), \\
    \frac{1}{2}{\rm Ch}_{4n+2}\left[\hat P\,(\bk_\parallel)\right] \equiv \nu^{d=8n+5}_{\rm class\ CI}\left[\mathcal{H}_{\rm bulk}\,(\bk)\right] \quad \left( \mathrm{mod}~2\right), \\
    \nu^{d-1=8n+6}_{\rm class\ AI}\left[\hat P^{(+)}\,(\bk_\parallel)\right]+\nu^{d-1=8n+6}_{\rm class\ AI}\left[\hat P^{(-)}\,(\bk_\parallel)\right] = \nu^{d=8n+7}_{\rm class\ BDI}\left[\mathcal{H}_{\rm bulk}\,(\bk)\right], \\
    \nu^{d-1=8n+7}_{\rm class\ AI}\left[\hat P^{(+)}\,(\bk_\parallel)\right]+\nu^{d-1=8n+7}_{\rm class\ AI}\left[\hat P^{(-)}\,(\bk_\parallel)\right] = \nu^{d=8n+8}_{\rm class\ BDI}\left[\mathcal{H}_{\rm bulk}\,(\bk)\right], \\
    \nu^{d-1=8n+7}_{\rm class\ AI}\left[\hat P\,(\bk_\parallel)\right] = \nu^{d=8n+8}_{\rm class\ D}\left[\mathcal{H}_{\rm bulk}\,(\bk)\right].
    \label{eq:BBC_d_8}
\end{align}
Here, the zeroth Chern number ${\rm Ch}_0\,[\hat P]$ is the rank of $\hat P$. 
These formulas~(\ref{eq:BBC_chiral})-(\ref{eq:BBC_d_8}) clarify the topology of the detached boundary states, leading to unique boundary responses and disorder effects.

\begin{figure}[H]
    \centering
    \includegraphics[scale=0.06]{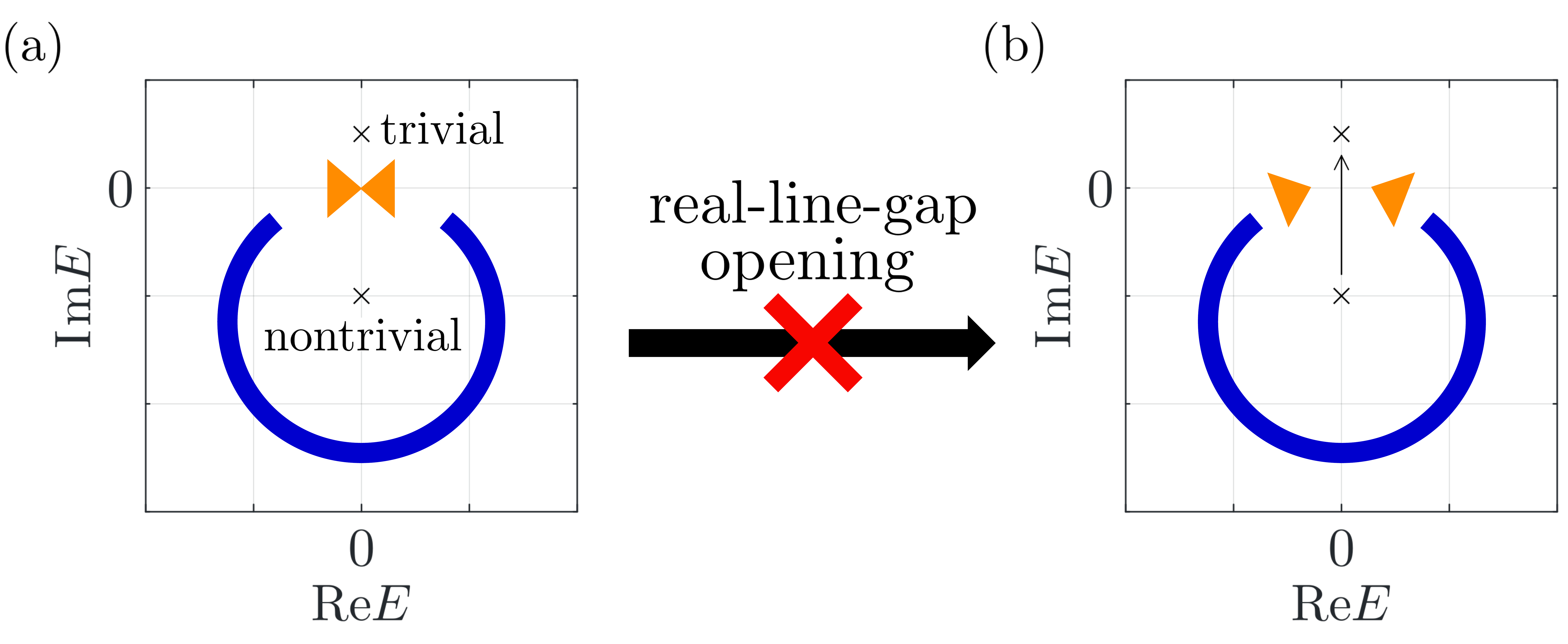}
    \caption{Prohibition against real-line-gap opening due to nontrivial point-gap topology.
    }
        \label{afig: real line gap opening}
\end{figure}

Meanwhile, the stability of the detached boundary states is ensured by real-line-gap (or equivalently, Hermitian) topology~\cite{Bessho-21}.
In fact, if a real line gap were open, point-gap topology could be trivialized by continuously moving the reference point far away, which would contradict the stability of point-gap topology (see Fig.~\ref{afig: real line gap opening}).
Consistently, all the extrinsic non-Hermitian topology reduces to imaginary-line-gap topology, as summarized in Table~\ref{tab:SFH_AZdag}.

\section{Intrinsic non-Hermitian topology}
    \label{asec: intrinsic}

\begin{table}[b]
\caption{Homomorphisms $f_r, f_i$ from line-gap to point-gap topology for the tenfold Altland-Zirnbauer$^{\dag}$ classification.}
\label{tab:SFH_AZdag}
\centering
{\scriptsize
$$
\begin{array}{ccccccccccccc}
\hline \hline
\mbox{Class} & \mbox{Gap} & d=0&d=1&d=2&d=3&d=4&d=5&d=6&d=7 \\
\hline
{\rm A}& {\rm L} \to {\rm P} &\Z \to 0&0 \to \Z&\Z \to 0&0 \to \Z&\Z \to 0&0 \to \Z&\Z \to 0&0 \to \Z\\ 
&&&&&&&&&\\ 
\hline
{\rm AIII}& {\rm L_r}\to {\rm P}&0 \to \Z&\Z \to 0&0 \to \Z&\Z \to 0&0 \to \Z&\Z \to 0&0 \to \Z&\Z \to 0\\
&&&&&&&&&\\ 
&{\rm L_i} \to {\rm P}&\Z\oplus\Z\to\Z&0\to 0&\Z\oplus\Z\to\Z&0\to 0&\Z\oplus\Z\to\Z&0\to 0&\Z\oplus\Z\to\Z&0\to 0\\ 
&&(n,m)\mapsto n-m&&(n,m)\mapsto n-m&&(n,m)\mapsto n-m&&(n,m)\mapsto n-m&\\ 
\hline
{\rm AI^\dag}& {\rm L} \to {\rm P}&\Z \to 0&0\to 0&0\to 0&0\to 2\Z&2\Z\to 0&0\to \Z_2&\Z_2\to \Z_2&\Z_2\to\Z\\ 
&&&&&&&&n\mapsto 0&\\
\hline
{\rm BDI^\dag}&{\rm L_r} \to {\rm P}&\Z_2 \to \Z&\Z \to 0&0\to 0&0\to 0&0\to 2\Z&2\Z\to 0&0\to \Z_2&\Z_2\to \Z_2\\ 
&&&&&&&&&n\mapsto 0\\ 
&{\rm L_i} \to {\rm P}&\Z\oplus \Z\to\Z&0\to 0&0\to 0&0\to 0&2\Z\oplus 2\Z\to 2\Z&0\to 0&\Z_2\oplus\Z_2\to\Z_2&\Z_2\oplus\Z_2\to\Z_2\\ 
&&(n,m)\mapsto n-m&&&&(n,m)\mapsto n-m&&(n,m)\mapsto n+m&(n,m)\mapsto n+m\\ 
\hline
{\rm D^\dag}&{\rm L_r} \to {\rm P}&\Z_2 \to \Z_2&\Z_2 \to \Z&\Z \to 0&0\to 0&0\to 0&0\to 2\Z&2\Z\to 0&0\to \Z_2\\ 
&&n \mapsto 0&&&&&&&\\
& {\rm L_i} \to {\rm P}&\Z \to \Z_2&0 \to \Z&0 \to 0&0 \to 0&2\Z\to 0&0\to 2\Z&\Z_2\to 0&\Z_2\to \Z_2 \\ 
&&n \mapsto n&&&&&&&n\mapsto n \\ 
\hline
{\rm DIII^\dag}& {\rm L_r} \to {\rm P}&0 \to \Z_2&\Z_2 \to \Z_2&\Z_2 \to \Z&\Z \to 0&0\to 0&0\to 0&0\to 2\Z&2\Z\to 0\\ 
&&&n \mapsto 0&&&&&&\\
&{\rm L_i} \to {\rm P}&\Z\to\Z_2&0\to\Z_2&\Z\to\Z&0\to 0&\Z\to 0&0\to 0&\Z\to 2\Z&0\to 0\\ 
&&n\mapsto n&&n\mapsto 2n&&&&n\mapsto n&\\ 
\hline
{\rm AII^\dag}& {\rm L} \to {\rm P}&2\Z \to 0&0 \to \Z_2&\Z_2 \to \Z_2&\Z_2 \to \Z&\Z \to 0&0\to 0&0\to 0&0\to 2\Z\\
&&&&n \mapsto 0&&&&&\\
\hline
{\rm CII^\dag}& {\rm L_r} \to {\rm P}&0 \to 2\Z&2\Z \to 0&0 \to \Z_2&\Z_2 \to \Z_2&\Z_2 \to \Z&\Z \to 0&0\to 0&0\to 0\\ 
&&&&&n \mapsto 0&&&&\\
&{\rm L_i} \to {\rm P}&2\Z\oplus 2\Z\to 2\Z&0\to 0&\Z_2\oplus\Z_2\to \Z_2&\Z_2\oplus \Z_2\to\Z_2&\Z\oplus \Z\to\Z&0\to 0&0\to 0&0\to 0\\ 
&&(n,m)\mapsto n-m&&(n,m)\mapsto n+m&(n,m)\mapsto n+m&(n,m)\mapsto n-m&&\\
\hline
{\rm C^\dag}& {\rm L_r} \to {\rm P}&0 \to 0&0 \to 2\Z&2\Z \to 0&0 \to \Z_2&\Z_2 \to \Z_2&\Z_2 \to \Z&\Z \to 0&0\to 0\\ 
&&&&&&n \mapsto 0&&&\\ 
& {\rm L_i} \to {\rm P}&2\Z \to 0&0\to 2\Z&\Z_2 \to 0&\Z_2 \to \Z_2&\Z \to \Z_2&0 \to \Z&0 \to 0&0 \to 0\\ 
&&&&&n \mapsto n&n \mapsto n&&&\\ 
\hline
{\rm CI^\dag}& {\rm L_r} \to {\rm P}&0 \to 0&0 \to 0&0 \to 2\Z&2\Z \to 0&0 \to \Z_2&\Z_2 \to \Z_2&\Z_2 \to \Z&\Z \to 0\\ 
&&&&&&&n \mapsto 0&&\\
& {\rm L_i} \to {\rm P}&\Z\to 0&0\to 0&\Z\to 2\Z&0\to 0&\Z\to\Z_2&0\to\Z_2&\Z\to\Z&0\to 0\\ 
&&&&n\mapsto n&&n\mapsto n&&n\mapsto 2n&\\
\hline \hline
\end{array}
$$
}
\end{table}

We classify intrinsic non-Hermitian topology, as summarized in Table~\ref{tab:SFH_AZdag} (see also Sec.~SIX in the Supplemental Material of Ref.~\cite{OKSS-20}, as well as Ref.~\cite{Shiozaki}).
As also discussed in the main text, if a line gap is open, a point gap is also open for a reference point chosen to be put on the reference line.
Thus, we can introduce a map from line-gapped topological phases to point-gapped topological phases for each spatial dimension and symmetry class.
Given a $d$-dimensional non-Hermitian Hamiltonian $H \left( \bm{k} \right)$, we consider the Hermitized Hamiltonian
\begin{equation}
    \tilde{H} \left( \bm{k} \right) \coloneqq \begin{pmatrix}
        0 & H \left( \bm{k} \right) \\
        H^{\dag} \left( \bm{k} \right) & 0
    \end{pmatrix}_{\sigma},
\end{equation}
which respects chiral symmetry $\Gamma H \left( \bm{k} \right) \Gamma^{-1} = - H \left( \bm{k} \right)$ with $\Gamma = \sigma_z$ by construction.
When the non-Hermitian Hamiltonian $H \left( \bm{k} \right)$ has a point gap, the Hermitized Hamiltonian $\tilde{H} \left( \bm{k} \right)$ also has a gap, and vice versa.
Thus, the topological classification of $H \left( \bm{k} \right)$ coincides with that of $\tilde{H} \left( \bm{k} \right)$, which is denoted by $K_{\rm P}$.
By contrast, the presence of a line gap imposes an additional constraint on $\tilde{H} \left( \bm{k} \right)$.
A non-Hermitian Hamiltonian $H \left( \bm{k} \right)$ with a real or an imaginary line gap can be continuously deformed to a Hermitian or an anti-Hermitian Hamiltonian while keeping the line gap and relevant symmetry, respectively~\cite{KSUS-19}.
Hermiticity or anti-Hermiticity of $H \left( \bm{k} \right)$ imposes another chiral symmetry with $\Gamma_{\rm r} = \sigma_y$ or $\Gamma_{\rm i} = \sigma_x$ on the Hermitized Hamiltonian $\tilde{H} \left( \bm{k} \right)$, respectively.
This additional chiral symmetry leads to the different topological classification denoted by $K_{\rm L_{r}}$ or $K_{\rm L_{i}}$ in the presence of a real or an imaginary line gap (and a point gap as well).
Forgetting $\Gamma_{\rm r}$ ($\Gamma_{\rm i}$) defines a homomorphism $f_{\rm r}: K_{\rm L_r} \to K_{\rm P}$ ($f_{\rm i}: K_{\rm L_i} \to K_{\rm P}$) from $K_{\rm L_r}$ ($K_{\rm L_i}$) to $K_{\rm P}$.
Owing to the dimensional isomorphism of the $K$-theory, it is sufficient to compute $f_{\rm r}$ or $f_{\rm i}$ in zero dimension to obtain those in arbitrary dimensions.
Table~\ref{tab:SFH_AZdag} summarizes these homomorphisms in the Altland-Zirnbauer$^{\dag}$ symmetry classes, which is relevant to this work.
See Refs.~\cite{OKSS-20, Shiozaki} for the homomorphisms in all the 38 symmetry classes.

If a point-gapped non-Hermitian Hamiltonian $H \left( \bm{k} \right)$ is included in the image of either homomorphism $f_{\rm r}$ or $f_{\rm i}$, it can be continuously deformed to a Hermitian or an anti-Hermitian Hamiltonian, implying that the topological nature is also attributed to the conventional Hermitian one.
Conversely, if $H \left( \bm{k} \right)$ is not included in the image of $f_{\rm r}$ or $f_{\rm i}$, its topological nature is intrinsic to non-Hermitian systems.
Such intrinsic non-Hermitian topology is captured by the quotient group $K_{\rm P}/\left( \mathrm{Im}\,f_{\rm r} \cup \mathrm{Im}\,f_{\rm i}\right)$, which is also summarized in Table~\ref{tab:SFH_AZdag}.
For example, point-gap topology in one-dimensional class A is intrinsic to non-Hermitian systems, while point-gap topology in two-dimensional class AIII is reducible to anti-Hermitian systems and hence of extrinsic nature.

\section{Reflection-invariant topological crystalline insulators}

Our formulation is also applicable to topological crystalline insulators.
As a prime example, we study two-dimensional topological insulators with commutative chiral and reflection symmetries~\cite{Chiu-13, Morimoto-13, Shiozaki-14}:
\begin{equation}
    \Gamma \mathcal{H} \left( k_x, k_y \right) \Gamma^{-1} = - \mathcal{H} \left( k_x, k_y \right), \quad \mathcal{R}_+ \mathcal{H} \left( k_x, k_y \right) \mathcal{R}_{+}^{-1} = \mathcal{H} \left( -k_x, k_y \right),
\end{equation}
with unitary matrices $\Gamma$ and  $\mathcal{R}_+$ satisfying 
\begin{equation}
\Gamma^2 = \mathcal{R}_+^2 = 1,\quad \left[ \Gamma, \mathcal{R}_+ \right] = 0.
\end{equation}

To apply our formalism, we focus on effective gapless Hamiltonians at boundaries. 
At an edge around a reflection-invariant corner, a gapless state 
\begin{equation}
\mathcal{H}_{{\rm AIII} + \mathcal{R}_+} \left( k_x \right) = k_x \sigma_x
\end{equation}
appears, where $k_x$ is a momentum along the edge of the two-dimensional system.
This Dirac model 
indeed respects the symmetries with $\Gamma = \sigma_z$ and $\mathcal{R}_+ = \sigma_z$.
Notably, this reflection-invariant gapless state exhibits no spectral flow and hence is detachable from the bulk, similarly to the Dirac surface state protected by chiral symmetry.
To confirm this, we couple it to a trivial band $\varepsilon \sigma_x$ ($\varepsilon \geq 0$) by
\begin{equation}
    \tilde{\mathcal{H}}_{{\rm AIII} + \mathcal{R}_+} \left( k_x \right) = \begin{pmatrix}
        k_x \sigma_x & v \sigma_+ \\
        v \sigma_- & \varepsilon \sigma_x
    \end{pmatrix},
\end{equation}
with the coupling strength $v \geq 0$.
This four-band model still respects chiral symmetry with $\Gamma = \sigma_z$ and reflection symmetry with $\mathcal{R}_+ = \mathrm{diag} \left( \sigma_z, 1 \right)$.
Similarly to the chiral-symmetric two-dimensional Dirac model studied in Sec.~\ref{asubsec: boundary 3D class AIII}, the four energy bands are obtained as
\begin{equation}
    \sqrt{2} E \left( k_x \right) = \pm \sqrt{ k^2_x + \varepsilon^2 + v^2 \pm \sqrt{\left( k^2_x - \varepsilon^2+ v^2 \right)^2 + 4\varepsilon^2v^2} },
\end{equation}
exhibiting a global gap in Eq.~(\ref{aeq: 3D AIII - global gap}).
Since gap-opening perturbations can further be added away from reflection-invariant corners, this also implies the emergence of corner states at zero energy~\cite{Langbehn-17}.

Corresponding non-Hermitian systems $H_{{\rm AIII} + \mathcal{R}_+} \left( k_x \right)$ respect chiral symmetry and commutative reflection symmetry:
\begin{equation}
    \Gamma H_{{\rm AIII} + \mathcal{R}_+}^{\dag} \left( k_x \right) \Gamma^{-1} = - H_{{\rm AIII} + \mathcal{R}_+} \left( k_x \right), \quad \mathcal{R}_+ H_{{\rm AIII} + \mathcal{R}_+} \left( k_x \right) \mathcal{R}_{+}^{-1} = H_{{\rm AIII} + \mathcal{R}_+} \left( -k_x \right), \quad \left[ \Gamma, \mathcal{R}_+ \right] = 0.
\end{equation}
Their point-gap topology is captured by the zeroth Chern numbers of the Hermitian matrix $\ii\,[ H_{{\rm AIII} + \mathcal{R}_+} \left( k_x \right) -E_{\rm P} ]\,\Gamma$ at reflection-invariant momenta.
In the presence of an imaginary line gap, by contrast, $\ii H_{{\rm AIII} + \mathcal{R}_+} \left( k_x \right)$ hosts two pairs of independent zeroth Chern numbers depending on $\Gamma = \pm 1$.
Consequently, the point-gap topology is of extrinsic nature, underlying the detachment of reflection-invariant corner states.


\end{document}